# Self-supervised learning for multiplexing super-resolution confocal microscopy

Qinglin Chen[1,†], Luwei Wang[1,†], Jia Li[1], Dan Shao[2], Xiaoyu Weng[1], Liwei Liu,[1], Dayong Jin[2,3,*] & Junle Qu[1,*]

[1]State Key Laboratory of Radio Frequency Heterogeneous Integration (Shenzhen University) & Key Laboratory of Optoelectronic Devices and Systems, College of Physics and Optoelectronic Engineering, Shenzhen University, Shenzhen 518060, China

[2]Institute for Biomedical Materials and Devices (IBMD), Faculty of Science, University of Technology Sydney, NSW 2007, Australia

[3]Zhejiang Provincial Engineering Research Center for Organelles Diagnostics and Therapy, Eastern Institute of Technology, Ningbo 315200, China

[†]These authors contributed equally: Qinglin Chen, Luwei Wang.

[*]Corresponding author: dayong.jin@eitech.edu.cn; jlqu@szu.edu.cn

Confocal microscopy has long been a cornerstone technique for visualizing complex interactions and processes within cellular structures. However, achieving super-resolution imaging of multiple organelles and their interactions simultaneously has remained a significant challenge. Here, we present a self-supervised learning approach to transform diffraction-limited, single-colour input images into multi-colour super-resolution outputs. Our approach eliminates the need for paired training data by utilizing a degradation model. By enhancing the resolution of confocal images and improving the identification and separation of cellular targets, this method bypasses the necessity for multi-wavelength excitation or parallel detection systems. Trained on an extensive dataset, the model effectively distinguishes and resolves multiple organelles with high fidelity, overcoming traditional imaging limitations. This technique requires no hardware modifications, making multi-colour super-resolution imaging accessible to any standard confocal microscope. We validated its performance by demonstrating two- and three-colour super-resolution imaging of both fixed and live cells. The technique offers a streamlined and data-efficient solution for multi-channel super-resolution microscopy, while opening new possibilities for investigating dynamic cellular processes with unprecedented clarity.

## Introduction

Fluorescence confocal microscopy has revolutionized cellular biology by enabling detailed visualization of specific subcellular compartments and organelles, thereby offering critical insights into the intricate mechanisms of life [1–3]. However, imaging multiple structures often necessitates different excitation and detection spectra, requiring multiple excitation/emission cycles that proportionally slow down imaging speed. To overcome this limitation, simultaneous excitation using multiple lasers and multi-spectral detection systems have been developed, allowing imaging of up to six organelle types and their interactions [4]. Despite these advances, the approach introduces challenges such as the increased phototoxicity from multiple laser excitation, which poses significant constraints for live cell imaging. Additionally, the inherent diffraction barrier further limits the resolution of fine organelle details, emphasizing the need for super-resolution imaging techniques [5,6].

The advent of super-resolution microscopy techniques, such as stimulated emission depletion (STED), structured illumination microscopy (SIM) and single-molecule localization microscopy (SMLM), has significantly advanced the field of subcellular imaging by overcoming the diffraction limit of light [7–11]. The growing need for simultaneous observation of multiple organelles has driven the development of methods capable of resolving spectral overlaps between fluorescent dyes. Current multi-colour super-resolution imaging approaches often rely on sequential imaging and spectral unmixing, achieved through combination of multiple lasers, detectors, filter-based fluorescence selections and colour compensation efforts [12]. For instance, multi-colour STED imaging has demonstrated three-colour super-resolution using a single depletion laser at 775 nm to simultaneously deplete fluorescence signals excited by three lasers at different wavelengths [13–18]. While effective, this strategy requires chromatic aberration correction across a wide spectral range, and the improvement in spatial resolution often comes at the expense of temporal resolution, which is critical for live-cell imaging.

To address these limitations, advanced methods like hyperSTED and phasor-STED have been developed. HyperSTED uses a single wavelength excitation laser, a single STED laser, and hyperspectral detection to simultaneously image up to four fluorescent markers [19]. Similarly, phasor-STED achieves five-channel imaging by combining fluorescence lifetime imaging (FLIM) with a single depletion laser and multi-wavelength excitation lasers [20]. These methods offer stability and the ability to simultaneously acquire multi-colour data, minimizing sequential imaging steps. However, they present practical challenges: Selecting a compatible panel of fluorescent probes remains complex, as spectral overlaps can limit the accuracy of signal separation. These techniques often necessitate specialized and sophisticated hardware, including multiple lasers and advanced optical setups. In regions where fluorescence channels overlap, incomplete or ambiguous data can arise, complicating quantitative analyses.

Deep learning has attracted considerable interest in fluorescence microscopy for its ability to overcome traditional imaging limitations and unlock new capabilities [21–26]. Virtual staining methods based on deep learning—used to transform label-free imaging modalities into virtually labeled images—have been successfully applied to cells and tissue specimens [27,28]. Despite their innovative nature, these methods remain limited by the optical diffraction barrier of the original imaging modality, making them insufficient for the precise visualization of intracellular structures and organelles [29,30]. Various deep learning-based approaches have been employed to surpass the optical diffraction limit, eliminating the need for complex optical systems [31–33]. These methods have demonstrated significant promise, yet challenges persist, particularly when applied to multi-colour imaging. Deep learning techniques often require sequential imaging or spectral unmixing algorithms to differentiate signals from multiple fluorescent dyes. While effective in static samples, these approaches are less suited for live-cell imaging, where temporal resolution is critical. Moreover, deep learning-based approaches for long-term live-cell imaging are limited in their ability to capture more than two or three colours simultaneously. This is partly due to challenges such as spectral crosstalk, photobleaching, and phototoxicity. Recently a deep learning-based channel unmixing method has been applied on wide-field and confocal microscopy, it utilizes emission distribution in channels to reveal different organelles components [34,35], while great amounts of spectral data are still required. A gap remains in utilizing these techniques for multi-colour imaging of multiple organelles and subcellular compartments in super-resolution microscopy.

The goal of the proposed work is to achieve multi-colour super-resolution imaging for intracellular interactome observation with simplified system configuration, synchronized acquisition and low photodamage. Existing deep learning techniques for fluorescence image enhancements often rely on paired training data, which makes obtaining high-quality imaging data very challenging and time-consuming. The dependence on paired data presents a major challenge in this work, as acquiring single-colour diffraction-limited and multi-colour super-resolution data simultaneously during live-cell dynamic processes is extremely difficult, while multi-staining in same organelle may introduce non-coincide regions in channels[36]. Moreover, multiple probes staining may introduce decreasing expression efficiency, yielding inaccurate mapping between input and ground truth (GT) [35]. In most applications of multi-channel super-resolution microscopy for intracellular observations, regardless of involved fluorescence dyes, different components are tagged and observed. Based on this fact, we assume in multi-colour fluorescence imaging, though components are imaged in different channels, their 2D information exists in the same axial vicinity. Enlightened by recent physics-driven methods for microscopy [33,37], we developed a self-supervised learning method to achieve multi-colour super-resolution mapping from single-channel diffraction-limited confocal microscopy.

Here, we present Deep-DSCM (deep learning for decoupling and super-resolution confocal microscopy) for multi-colour and super-resolution imaging using a single-channel confocal system. By introducing the degradation model, we are able to simulate low-resolution input images closely resembling real-world scenarios, significantly reducing the difficulty and cost of data acquisition. There are several synthetic paired supervised training methods, which have real-world GT and artificial input, combined as training pairs [38,39], however, our method does not require real-world input and GT, they are both derived from STED images of individual organelles, the inputs and outputs are originated from identical samples, making the proposed method fall in a self-supervised manner. Utilizing deep neural networks trained on diverse datasets of cellular morphologies, our method achieves effective segmentation of two- and three-colour organelle structures, when the dyes exhibit pronounced spectral overlaps. In summary, Deep-DSCM successfully circumvents the reliance on complex hardware and paired training data in traditional multi-colour super-resolution imaging methods by integrating deep learning techniques and an innovative degradation model. This approach offers an economical, efficient, and flexible solution for multi-colour super-resolution imaging within cellular structures.

## Results

### Principle of Deep-DSCM

Figure 1 presents the scheme of Deep-DSCM based on a semi-supervised approach assisted by a degradation model. The data required for the method is only individual STED images from different samples. To obtain the paired images from both single-colour low-resolution and multi-colour HR images, the method applies a degradation model to simulate realistic single-channel confocal images by blurring, compositing and introducing noise to multiple STED images. Starts with STED images, multiple STED images of organelles of sole kinds were randomly selected and stacked to serve as GT. First GT images were scaled to a unified range and multiplied with a organelle-specific factor to ensure close intensity among organelles. Next STED images were blurred with modulated PSF, which is calculated given the resolution of both STED images and target low-resolution images. STED imaging can have different super-resolution effects on different settings (wavelength, photobleaching, depletion power, etc.). In this paper the resolution of STED images was estimated by single rFRC mapping method [40], given in Figure S1. Then the images are straightforwardly composited into one-channel image, simulating single-channel imaging. To approach the noise situation in real-world imaging, a mixture of Gaussian and Poisson noise and frame averaging are imposed on the image, finally providing a single-channel, low-resolution, noise-contaminated image, which serves as input for the network. (see Figure 1, Methods, Supplementary Note 2 and Figure S2).

It is noteworthy that each channel of multi-colour STED images is acquired from different single-stained samples, which maximizes the staining efficiency and preserves the completeness of organelle structures. During training, the network processes input images and generates predictions that are compared to ground truth (GT) images. To demonstrate the effectiveness of the degradation process, we conducted super-resolution experiments for models trained with real-world / pseudo dataset for comparison (See in Supplementary Fig 6). This approach can eliminate the need in matching the colocalized datasets of input and GT images, which is labor-intensive, thereby reducing the risk of image misalignment. To indicate the proposed method is not model-reliant, we employed a single U-Net architecture as generator. During training MSE loss, VGG-19 based feature loss, SSIM loss, and U-Net based GAN loss were applied for enabling the network to preserve high frequency features (see Methods, Supplementary Note 1 and Figure S1).

To discuss the edge of our method's ability, we performed an experiment on pure simulated data, with lines and ellipses, which varies to be identical (See in supplementary Note 3 and Figure S3). The results show that the method is capable to generate detangled super-resolved images even in high similarity (SSIM=0.994), while the network failed in lines-lines (identical) compositions. However, in practical fluorescence imaging, detected pixel value varies due to many factors, causing different intensities among stained components, and this intensity difference introduces additional clues for the network, making the method still applicable in high similarity conditions (Figure S3c, row 4).

**Evaluation of synthetic data for multi-colour super-resolution imaging of organelles**

We first evaluated the performance of Deep-DSCM in decoupling and resolution enhancement using synthetic images of subcellular structures without artificial noise. High-resolution STED images of nuclear pore complexes (NPCs) stained with STAR RED, inner mitochondrial membrane (IMM) stained with PK Mito orange [41], and the microtubules stained with STAR RED were acquired using the STEDYCON system and the Facility line system [42]. Both synthetic single-colour low-resolution noise-free images and multi-colour super-resolution images of two or three organelles were generated by degradation model as training data (see Methods).

Figure 2 shows that our approach effectively decouples each organelle and enhances their resolution and fidelity from the input images, closely matching GT data. In the single-colour images in the first column of Figure 2a, organelle details appear indistinct and ambiguous. In contrast, Deep-DSCM successfully decouples and super-resolves the majority of structures, including the distinguishing features of adjacent organelles. For images with less structural complexity, overlapping structures from two types of organelles can also be resolved with intact organelles' profiles. For example, in regions where microtubules and IMM overlap, the densely packed microtubules and IMM's fine cristae structures are separated and resolved well, preserving their detailed and intricate features. To illustrate the reconstruction results, we provide the corresponding resolution-scaled Pearson coefficient (RSP), resolution-scaled error (RSE), and RSE maps [43] in

Figure 2a. These metrics compare diffraction-limited images to their super-resolution counterparts to highlight super-resolution defects. The Deep-DSCM reconstructions exhibit low reconstruction errors for two organelles. In conclusion, the trained network is able to differentiate organelles while a slight degree of error for three organelles appears as the structural complexity increases, shown in magnifications in Figure 2a. As Figure 2a shows (rightmost plots), the reconstruction errors mainly occurred in the overlapped region, in which the structure information is buried. While the networks give good classification on organelles in discrete regions.

To quantitatively assess the overall performance of Deep-DSCM, we conducted assessment on 200 samples for each combination. We used the three additional metrics of mean absolute error (MAE), structural similarity index matrix (SSIM), peak signal-to-noise ratio (PSNR), and rolling Fourier ring correlation (rFRC) analysis [40]. These metrics evaluate the quality of the decoupled super-resolved images. MAE measures pixel-level data fidelity by calculating the $l_1$ norm of the difference between the reconstructed images and GT data. SSIM is an image quality metric that measures the perceptual similarity between two images by comparing their luminance, contrast, and structural information. PSNR reflects the reconstruction quality of the output images, while rFRC evaluates uncertainties in the super-resolution reconstructions and provides a detailed regional map for examination. As shown in Figure 2b, Deep-DSCM exhibits high fidelity based on MAE, SSIM and PSNR metrics, although the performance slightly declines when three types of organelles were involved. Compared to GT images with a resolution of approximately 85 nm, the output images from Deep-DSCM maintain super resolutions below 100 nm for all the organelle types and their details. The corresponding resolution maps for these results are provided in Figure S4.

Supervised learning methods require tons of data to perform well on validation, however, in super-resolution fluorescence microscopy, huge, paired dataset of super-resolved images does not exist. In our proposed training method, the network is free of both paired single-channel confocal images and multi-channel super-resolution images. Instead, the training method generates great amount of data pairs by randomly selecting organelle images and compositing them during the training process. Benefited by that, the model is data-efficient for experimental acquisition, especially on small amounts of data. In Supplementary Note 5 and Figure S5, we performed a comparison on a tiny raw dataset which only includes 20 images of microtubules and mitochondria STED images as raw data. We compared the training on three different dataset configurations, the only difference was the number of generated input and GT images: we generated 20 and 1000 sets as prepared data, while training data were loaded and trained, and another set was kept using the proposed method. The results are shown in Fig S5, based on random selection process in degradation model, the training process is much more stable, and the trained network can still perform well though the amount of data is limited by augmenting the data inside the model.

**Deep-DSCM multi-colour super-resolution imaging of confocal images**
We next applied Deep-DSCM to distinguish and super-resolve multiple organelles from single-colour diffraction-limited confocal images as real-world situation, other three types of organelle structure were involved (microtubules, mitochondria and lysosomes). To demonstrate the generalizabilityof our model on other imaging platforms, all the real-world data (single-channel confocal images) were samples from other systems (Facility Line, Abberior, SP8, Leica, Home-built). We first demonstrate evaluation of real dataset on single-structure images, shown in Figure S6, microtubules, mitochondria and NPCs fixed samples are applied for single-colour super-resolution experiments. The network trained on synthetic data performed well on the real-world confocal data.

As shown in Figure 3a, we first acquired microtubules-mitochondria images of prepared samples on SP8 system (Leica), in which deep red (640 nm, input) channel was stained with microtubules and mitochondria, in red (561 nm, GT) channel microtubules were stained, and in green channel (488 nm, GT) mitochondria were stained. We conducted confocal imaging in each channel since STED imaging performance may vary due to the spectral difference. The results are shown in Figure 3a, despite the model is trained on fully synthetic data, it still has good ability to detangle and super-resolve the organelles. We used resolution scaled metrics to estimate confocal GT and super-resolved network inference images. The average RSP and RSE were 0.871 and 11.977, even better than its performance on synthetic dataset (0.850, 13.742). To investigate how different parameters of the degradation model influence its performance on real-world microscopy data, we conducted additional comparisons presented in Supplementary Figure S7. The results indicate that variations in parameters lead to slight differences in performance, the trained network has the generalizability for small domain offset.

Other additional confocal images in Figure 3 were obtained using our custom-built confocal system [44,45], where microtubules, mitochondria, and lysosomes in live HeLa cells were stained with Tubulin Tracker™ Deep Red (Thermofisher), MitoTracker™ Deep Red (ThermoFisher), and LysoBriter™ NIR (AAT Bioquest), respectively. Evaluation results for these organelles on a synthetic dataset were presented in Figure S8 and Figure S9. In all the three types of stained combinations, shown in Figure 3b, the closely spaced organelles cannot be distinguished from the confocal microscopy images (first column) due to the diffraction limit and noise inherent to scanning microscopy. In contrast, our Deep-DSCM approach can successfully decouple them (second to fourth columns) with enhanced resolutions. This means that only single-colour confocal images of multiple organelles are needed to produce super-resolution images, suggesting distinct advantages compared with existing methods that either require imaging one organelle per channel or only produce the merged images without separating individual organelles [13–18].

The results of other combinations are shown in Figure 3b. As the magnified results shown in the yellow boxes, the super-resolution images of decoupled organelles display sharper and more complete structures compared to the raw confocal images. For example, our method effectively distinguishes lysosomes from mitochondria (Figure 3b, first row) and detangles mixed microtubules and lysosomes information (Figure 3b, second row), which appear blurred and indistinguishable in the diffraction-limited single-colour confocal images. This is further validated by the intensity profiles along the white dashed lines, where diffraction artifacts and noise between adjacent structures in the confocal images are significantly reduced. Moreover, our Deep-DSCM predictions effectively preserve key cellular information, thresholded versions of the aforementioned results were also presented in Figure S10 to support this conclusion. It is noteworthy that these real-world datasets were obtained across different confocal systems (Supplementary table S3), suggesting the adaptability of Deep-DSCM approach.

To quantify the resolution improvement achieved by Deep-DSCM, we repeated the analysis of 21 images for microtubules-mitochondria, 55 images for mitochondria-lysosomes, 36 images for microtubules-lysosomes, and 24 images for microtubules-mitochondria-lysosomes, and measured the resolutions of the decoupled images using single-frame rFRC mapping. As summarized in Figure S11, the resolution of the network's outputs is approximately 95 nm. Notably, the resolution measurement tends to be more statistically stable in microtubules-mitochondria (fixed sample), due to its higher signal-to-noise ratio than others, indicating better imaging condition can bring a better reconstruction results of the proposed method. The degradation model can be adapted to different imaging conditions by adjusting parameters for specific organelles, convolution PSF, and noise levels (see Figure S2). Additionally, the degradation model and adversarial loss in training gives Deep-DSCM strong adaptation to varying imaging scenarios.

**Deep-DSCM for organelle interactions in living cells**

We further showcased Deep-DSCM in live cell imaging to study the interactions between subcellular structures, which are essential for understanding physiological and pathological processes [46]. Time-lapse imaging of mitochondria and lysosomes in live HeLa cells was conducted using a custom-built confocal microscopy system. Due to the fact that the applied fluorescence dyes are susceptible to photobleaching and the interactions between mitochondria and lysosomes usually take a long time, the imaging was performed with a 30-second interval between frames. As shown in Figure 4a-c, the diffraction-limited single-colour confocal images were transformed by Deep-DSCM to reveal and track the dynamics of mitochondria and lysosomes and their interactions (Visualizations 1,2). Here, three processes of mitochondria-lysosomes autophagy, mitochondria-lysosomes contact (MLC) and mitochondria-lysosomes fission [47,48], vital for maintaining cellular homeostasis in eukaryotic cells [49], are highlighted in dashed boxes (Figure 4a). Magnified time-lapse images of these dynamic interactions at different time points are displayed in Figures 4b and 4c, as well as in Visualizations 1 and 2, respectively. Deep-DSCM can visualize the dynamic process of mitophagy and track changes in both organelles during mitochondrial degradation. Figures 4b illustrate a mitophagy process, where mitochondrion (green) is engulfed by a lysosome (magenta). In contrast, conventional single-channel confocal microscopy struggles to distinguish between mitochondria and lysosomes and fails in simultaneously revealing their interactions. Figure 4c shows both the MLC and fission process in one FOV, indicating that Deep-DSCM can support long-term imaging for intracellular process. Our network captures this 10-minute dynamic process (Visualization 2), showing the interaction from initial contact to eventual separation, and the fission process of mitochondria by lysosomes. This level of details is undetectable with traditional single channel confocal microscopy.

We also applied Deep-DSCM to super-resolve microtubules and lysosomes to observe their dynamic interactions (Figure 4d-f, Visualizations 3,4). A total of 143 frames of raw images were acquired at three-second intervals. As microtubules function as highways for cargo delivery [50,51], super-resolution microscopy can be used to monitor vesicle behaviours at microtubule crossings during the cargo transport process [51]. However, earlier methods either lacked high-resolution microtubule structures or required the registration of vesicle trajectories along microtubules in sequential STORM images. Here, leveraging the high resolution and single-colour imaging capabilities of Deep-DSCM, we can simultaneously monitor the dynamics of lysosomes and microtubules, allowing us to investigate how microtubules influence and regulate lysosomal movement.

Figure 4e,f present two types of lysosome movements on microtubule, i.e. along a single microtubule followed by a switch onto an intersecting microtubule (Figure 4e, Visualization 3), and along a microtubule back-and-forth before reversing after passing a microtubule intersection (marked by the red orange arrow, Figure 4f, Visualization 4). These distinct behaviours of lysosomes at microtubules' crossings are essential for studying intracellular transport [51,52] and can be subtle and undetectable from the low spatial resolution confocal images. Therefore, Deep-DSCM provides

an effective means to visualize the motion of lysosomes and microtubules, enabling dynamic tracking of their interactions with high resolution. Deep-DSCM clarifies the morphology of various organelles, helps determining their relative positions and movement trajectories, and facilitates the study of interactions among different organelles in dynamic process.

Furthermore, unlike STED super-resolution setups, the Deep-DSCM-assisted confocal approach achieves the results shown in Figures 3,4 without requiring high-power depletion lasers. This eliminates key concerns associated with photo-bleaching and phototoxicity, making the technique far more suitable for long-term live cell imaging applications.

**Discussion and Conclusion**

In this paper, we present a self-supervised method to achieve multi-colour super-resolution imaging of organelles based on a single-channel confocal microscopy system. This approach eliminates the need for complex hardware setups, including multiple lasers, detectors, and wavelength-selective devices, alleviates the photobleaching and phototoxicity concerns in STED super-resolution imaging, and bypasses the spectral cross talks in multi-colour imaging experiments. The key lies in the degradation model enabled self-supervised deep learning approach. The degradation model simulates the loss of channel number, image resolution, and signal noise, significantly lessening the burden of sample preparation and data acquisition. Random selection from different organelles greatly improves data efficiency when only a few images are available. By integrating adjustable blurring and noise parameters, the self-supervised training strategy demonstrates strong adaptability to real-world data on different imaging platforms, showcasing the network's robustness.

Though Deep-DSCM shows strong reconstruction quality when dealing with distinct organelles, reconstruction accuracy tends to decline when the organelles appear similar or when the number of organelles increases. Our method may struggle to assign overlapping structures to the correct channels due to local reconstruction errors, which are common in deep learning approaches. The current degradation model in 2D space assumes that organelles in all channels are at the optimal focal plane, even though some organelles may arrange at varying depths. This discrepancy may contribute to reconstruction failures, particularly in areas where structures overlap. Future work may integrate depth simulation into the degradation model.

As Deep-DSCM is based on the most commonly used confocal setup, our method has the potential to be extended to a wider range of fluorescence microscopy techniques, such as SIM, STORM, and multi-photon imaging, as well as spectral and phasor-based methods, their intrinsic information can further boost the reconstruction accuracy. Deep-DSCM will enable the integration of complementary imaging modalities and analytical techniques to reveal a comprehensive structural view of intracellular compartments and organelles' interactome in a cell.

**Materials and methods**

**Cell culture**

The Biologics Standards-Cercopithecus-1 (BSC-1) cell line was obtained from Pricella Life Technology Co., Ltd. BSC-1 cells were cultured in Dulbecco's modified Eagle's medium (DMEM) (Invitrogen, #11965-118) supplemented with 10% foetal bovine serum (FBS) (Gibco, #16010-159). To reduce the risk of bacterial contamination, 100 μg/ml penicillin and streptomycin (Invitrogen, #15140122) were added to the medium. The cells were maintained under standard cell culture conditions at 37°C in a humidified atmosphere with 5% $CO_2$.

When the cells reached over 80% confluence, they were passaged. For cell passage, the cells were washed three times with prewarmed PBS (Life Technologies, #14190500BT) and then digested with 25% trypsin (Gibco, #25200-056) for 30 s. Once the cells became significantly rounded and the cell spacing increased, digestion was terminated by adding media. The cell layer was gently disrupted to disperse the cells, which were then collected, centrifuged to remove trypsin, resuspended in fresh medium, and split into new culture bottles according to the passage ratio. Finally, the flasks were supplemented with the appropriate medium and returned to the incubator for continued culture. The BSC-1 cell line was tested for mycoplasma contamination using the MycoAlert detection kit (Lonza), and all tests confirmed the absence of contamination.

HeLa and MDA cells were also cultured in DMEM (#11965118, Thermo Fisher Scientific) supplemented with 10% foetal bovine serum (#26140079, Thermo Fisher Scientific) and a penicillin–streptomycin solution (100 units of penicillin and 100 μg/mL streptomycin in 0.85% saline; GenClone). These cells were incubated under the same conditions at 37°C in a 5% $CO_2$ humidified atmosphere.

**Sample preparation**

For the fixed cell samples, we prepared nuclear pore complexes (NPCs), microtubules, mitochondria, their dual-stained samples, and the plasma membrane. For NPCs, we purchased standard slice samples from Abberior, with NPCs stained with STAR RED fluorescence dyes for deep red excitation. For microtubules, mitochondria, and their dual-stained samples, BSC-1 cells were plated on #1.5 glass-bottom dishes 48 h prior to sample preparation. These cells were grown on 35 mm, #1.5 glass coverslips (SunBloss™, STGBD-035-1), with glass-bottom dishes pretreated with fibronectin (SunBloss™, HXAR-01) for 1 h at 37°C to increase adhesion. On the day of sample preparation, the cell density was maintained at approximately 50%-70%.

The cells were fixed at 37°C in prewarmed fixation buffer containing 4% paraformaldehyde and 0.1% glutaraldehyde (SunBloss™, HXKx01) in PBS for 10 min. After fixation, the samples were washed three times with PBS. To reduce background fluorescence, the cells were incubated in 2 ml of 0.1% $NaBH_4$ solution (SunBloss™, HXIK023) in PBS for 7 min with gentle shaking (< 1 Hz). The samples were then washed three times with 2 ml of PBS. Next, the cells were incubated for 30 min at 37°C in a blocking solution of PBS containing 5% BSA and 0.5% Triton X-100 (SunBloss™, HXKx02). The antibodies were diluted in the same blocking solution.

The samples were incubated for 40 min at 25°C with the appropriate dilutions of the following primary antibodies: beta-tubulin (DSHB-E7) and Tom20 (ABclonal, A19403). Afterwards, the samples were washed three times with 2 ml of PBS, with each wash lasting 5 min. The secondary antibodies (goat anti-mouse IgG, STRED-1001-500UG, and goat anti-rabbit IgG, STRED-1002-500UG, Abberior) were then applied for 60 min at 25°C. During this step, the samples were protected from light. Following three additional PBS washes, the cells were fixed for 10 min in a post-fixation buffer.

For live cell samples, a total of $2 \times 10^5$ cells were seeded on a glass-bottom microwell dish and incubated with 1 ml of DMEM supplemented with 10% FBS for 24 h. After overnight incubation, the cells were washed three times with PBS. For triple-stained samples of microtubules, mitochondria, and lysosomes, the cells were incubated in a 5% $CO_2$ atmosphere at 37°C for 15 min after 100 nM LysoBrite™ NIR (AAT Bioquest) diluted in DMEM (without 10% FBS) was added. The cells were washed 3 times with PBS, followed by incubation with 200 nM MitoTracker™ Deep Red FM (Thermo Fisher) for 30 min. The supernatant was then discarded. After treatment, the cells were washed 3 times in wash buffer and cultured in fresh medium for observation under a microscope.

For single-stained samples of the inner mitochondrial membrane and lysosomes, dual-stained samples of microtubules–lysosomes, mitochondria–lysosomes, and triple-stained samples of microtubules–mitochondria–lysosomes, the detailed preparation protocols and staining kit information are provided in Table S1.

**Raw data acquisition**

Experimental data for training both fixed and live cell samples were obtained using the STEDYCON system (Abberior), which is built on an inverted microscope (Olympus) equipped with a 100× 1.45 oil objective (HCX PL APO, 100×/1.40-0.9 OIL, Leica, Germany). Excitation light was provided by 561 nm and 640 nm lasers, whereas a 775 nm laser was used for depletion. Fixed-cell data for testing were collected using the SP8 system (Leica) with a 638 nm laser for excitation. Live-cell test data were acquired using our custom-built system, which utilized a 635 nm laser as the excitation source.

For each organelle, we prepared samples labelled with a single type of fluorescent dye to minimize potential channel cross-talk and ensure optimal image resolution, STED images were denoised using ImageJ's built-in denoising function. To create spectrally challenging samples for the testing dataset, we prepared multi-stained samples using spectrally overlapping dyes. Details of the setups used for the training and testing datasets are provided in Tables S2 and S3. To reduce the impact of noise, we utilized Huygens Essential software to apply deconvolution to images of the inner mitochondrial membrane.

**Network architecture**

Inspired by U-Net [53] and the application of a U-Net-based generative adversarial network (U-Net GAN) [32,54], we designed the Deep-DSCM system using a generative adversarial network (GAN) framework. The framework is illustrated in Figure S1a. Deep-DSCM comprises two components: a generative model and a discriminative model. Initially, a single-colour low-resolution image containing multiple organelles is fed into the generative model, which produces high-resolution images for each organelle. These generated images, alongside the corresponding ground truth (STED images of each organelle), form data pairs used as inputs and GT for the discriminative model. The discriminative model compares these pairs and assesses the pixelwise probability of the generated high-resolution images matching the STED images across both the original and downsampled scales. This training process continues iteratively until the predefined maximum number of epochs is reached. Based on our experience, the generator and discriminator generally achieve Nash equilibrium at this point [55].

The generator in Deep-DSCM adopts a classical U-Net structure, which enables it to suppress irrelevant regions while emphasizing important structures of varying shapes and sizes. This results in enhanced prediction performance across diverse datasets, as shown in Figure S1b. The encoder extracts spatial features from the input images, and the decoder reconstructs these features to generate super-resolution images for each organelle. Skip connections were established by concatenating the corresponding levels of the encoder and decoder. Each ConvBlock contains a Conv(3k1s1p) operation, which is a convolution with a 3 × 3 kernel, stride of 1, and padding of 1. This is followed by batch normalization to stabilize training and mitigate internal covariate shifts [56]. Additionally, we replaced rectified linear units (ReLU) with LeakyReLU [57,58] to reduce the likelihood of "dead neurons" and ensure gradient flow for negative inputs, improving convergence

stability:

$$LeakyReLU(X) = \begin{cases} x, & if\ x \geq 0 \\ x/a, & if\ x < 0 \end{cases} \qquad (1)$$

In the decoder, we alleviated the checkerboard artifacts commonly associated with transposed convolution [59] by first applying nearest neighbour interpolation for spatial upsampling by a factor of two. A subsequent 1 × 1 convolution layer then reduces the feature channels by half. Finally, an output block maps the resulting 64 channels into $N$ channels, corresponding to monochrome grayscale high-resolution images of $N$ organelles ($N$ equals 2 or 3).

The discriminator structure, detailed in Figure S1c, is a U-Net-based design inspired by single-frame super-resolution microscopy [32]. This architecture has been shown to yield richer per-pixel feedback for the generator, thereby improving the overall quality of generated images [32,54]. The encoder in the discriminator gradually downsamples the input, capturing the global image context, whereas the decoder performs progressive upsampling to enable pixelwise evaluations. Unlike standard classification-based discriminators, which provide only global decisions [60], this encoder–decoder framework retains both global and local data representations. Notably, we did not include a sigmoid activation function in the discriminator, as the BCEWithLogitsLoss() function used in this work already incorporates it.

**Loss function**

The discriminator loss, $L_D(X, Y)$, consists of two components: the encoder loss $L_{enc}(X, Y)$ and the decoder loss $L_{enc}(X, Y)$. These are derived from the per-pixel decisions made by the encoder module and decoder module of the discriminative algorithm [32,54]. Specifically, the input ground truth image $Y$ or the super-resolution image $G(X)$ generated by the generator is first convolved by the encoder to make a per-pixel decision $[D_{enc}(\cdot)]_{ij}$ on whether each pixel is real or generated. The decoder then progressively upsamples these encoded outputs to calculate the per-pixel decision $[D_{dec}(\cdot)]_{ij}$. The combined loss function is given as follows:

$$L_D(X,Y) = L_{enc}(X,Y) + L_{dec}(X,Y)$$

$$= -E\{\sum_{i,j} log[D_{enc}(Y)]_{i,j}\} - E\left\{\sum_{i,j} log\left(1 - [D_{enc}(G(X))]_{i,j}\right)\right\}$$

$$-E\{\sum_{i,j} log[D_{dec}(Y)]_{i,j}\} - E\left\{\sum_{i,j} log\left(1 - [D_{dec}(G(X))]_{i,j}\right)\right\} \qquad (2)$$

where $E\{\cdot\}$ represents taking the average for all the pixels in the image.

Meanwhile the generator loss $L_G(X)$ is written as:

$$L_G(X) = L_{enc}(G(X)) + L_{dec}(G(X))$$

$$= -E\left\{\sum_{i,j} log[D_{enc}(G(X))]_{i,j}\right\} - E\left\{\sum_{i,j} log[D_{dec}(G(X))]_{i,j}\right\} \qquad (3)$$

For the generative model, the loss function combines adversarial loss $L_{adv}(X, Y)$ and mean-square error (MSE) loss $L_{MSE}(X, Y)$. The adversarial loss is defined as follows:

$$L_{adv}(X,Y) = L_D(X,Y) + L_G(X) \qquad (4)$$

while the MSE loss computes the $l_2$ norm of the difference between the generator's prediction, $G(X)$, and the corresponding ground truth, $Y$:

$$L_{MSE}(X,Y) = \frac{1}{w \times h} \sum_{i=1}^{w} \sum_{j=1}^{h} (G(X)_{i,j} - Y_{i,j})^2 \qquad (4)$$

where $w$ and $h$ are the width and height of the images, respectively.

Furthermore, the VGG-19 based feature loss and SSIM loss are integrated into the generator loss, the feature loss is written as:

$$L_{fea}(X,Y) = L_{MSE}(Fea(X), Fea(Y)) \qquad (5)$$

*Fea* represents the feature extracted by the pretrained VGG-19 model. The SSIM loss is:

$$L_{SSIM}(X,Y) = 1 - SSIM(X,Y) \qquad (6)$$

The total generator loss, $L_G(X, Y)$, is expressed as follows:

$$L_G(X,Y) = \alpha \cdot L_{MSE}(X,Y) + \beta \cdot L_{SSIM}(X,Y) + \gamma \cdot L_{fea}(X,Y) + \delta \cdot L_{adv}(X,Y) \qquad (7)$$

where the constant factors balance the contributions of the respective terms. Empirical experiments were used to determine the optimal values of $\alpha = 1$, $\beta = 0.01$, $\gamma = 1$, and $\delta = 1$ for this work.

**Degradation model**

We trained the network exclusively in self-supervised manner, as it is extremely difficult to acquire a sufficient volume of real single-colour low-resolution images of multiple organelles along with their corresponding high-resolution images for training, particularly high-quality, well-registered experimental data pairs. To address this, we developed a degradation model that only requires super-resolution STED images of each organelle. These images were degraded to simulate low-resolution data, allowing us to generate well-registered image pairs conveniently and train the network with various organelles efficiently.

The details of degradation model are illustrated in Fig. S2. The degradation pipeline begins with a raw dataset comprising single-channel STED images of various organelles. For each organelle class, images are randomly sampled and normalized, then scaled by a constant to set organelle brightness across different samples, multi-channel super-resolution images are obtained as GT.

Based on the point spread function (PSF) resolution degeneration method of Ref. [22]. Here, the STED image $I_s$ and the confocal image $I_c$ are modelled as the convolution of the object *obj* with the point spread function (PSF) of their respective imaging systems.

$$I_s = obj \otimes sPSF$$

$$I_c = obj \otimes cPSF \tag{8}$$

where $\otimes$ denotes convolution and *sPSF* and *cPSF* represent the PSFs for the STED and confocal imaging systems, respectively. By applying a Fourier transform $FT(\cdot)$ to both equations and dividing them to eliminate the object's spectrum, we derive the following relationship:

$$FT(I_c) = FT(I_s) \cdot \frac{FT(cPSF)}{FT(sPSF)} \tag{9}$$

Using an inverse Fourier transform $IFT(\cdot)$ on the equation above, we generate a synthetic confocal image, enabling us to represent the low-resolution data on the basis of the high-resolution STED image. The *sPSF* of STED images $I_s$, is first generated, using the measured resolution of STED images then generate the *cPSF* of confocal systems by a manually given target resolution. Then, a modulation point spread function (*mPSF*) is computed to blur the STED images. Because the resolution of organelle STED images varies with imaging conditions, an *mPSF* is generated individually for each organelle, producing multi-channel, low-resolution images.

The degradation model's second step involves compositing low-resolution images of different organelles into a single-colour image.

In the final step, the synthetic image is corrupted with Poisson noise and Gaussian noise to more closely mimic experimental low-resolution images, frame averaging is applied for different confocal imaging settings. Notably, all model parameters can be adjusted to account for varying imaging conditions.

**Training**

For training, we collected STED images across six organelle types: 104 frames for nuclear pore complexes (NPCs), 115 frames for the inner mitochondrial membrane, 224 frames for microtubules, 196 frames for mitochondria, and 70 frames for lysosomes, we additionally imposed Huygens deconvolution to images of inner mitochondrial membrane to sharpen the images. The dataset was split, with 90% allocated for training and 10% allocated for validation. To further diversify the training data and improve the network's robustness, we applied data augmentation techniques, including random cropping and flipping. The training patch size was set to 384 × 384 pixels, and the batch size was set to 1, as this configuration has been shown to improve network performance in image generation tasks [61].

The training process for Deep-DSCM alternates between updating the generative model with the current discriminative model and updating the generator while keeping the discriminator fixed. We used the trained model at 1000 epochs (300 iterations in each epoch) as the final model for testing, which was sufficient for all the organelle types in our experiments. Both the generative model and discriminative model were optimized using the Adam optimizer [62], with initial learning rates of 0.00001 and 0.0000001, respectively. To accelerate the training process, we employed the CosineAnnealLR learning rate scheduler [63].

The framework was implemented via PyTorch (version 2.0.1) and Python (version 3.9.17) in Microsoft Windows 10. Training and inference were carried out on a workstation equipped with four NVIDIA GeForce RTX3090 GPUs and an Intel Xeon(R) Gold 6154 CPU @3.0 GHz. The inference time for an image of 512 × 512 pixels was approximately 6 ms.

**Evaluation metrics**

In this study, we evaluated the performance of our model using quality metrics such as the MAE, SSIM, PSNR, resolution-scaled analysis [40], and single-frame rFRC mapping [43]. The MAE metric is calculated using the following equation:

$$MAE(Y, \hat{Y}) = \frac{1}{w \times h} \sum_{j=1}^{w} \sum_{i=1}^{h} |\hat{Y}(i,j) - Y(i,j)| \tag{10}$$

where the width and height of the network output $\hat{Y}$ and the ground-truth image $Y$ are represented by $w$ and $h$, respectively.

The SSIM metric is:

$$SSIM(Y, \hat{Y}) = \frac{(2\mu_Y \mu_{\hat{Y}} + C_1)(2\sigma_{Y\hat{Y}} + C_2)}{(\mu_Y^2 + \mu_{\hat{Y}}^2 + C_1)(\sigma_Y^2 + \sigma_{\hat{Y}}^2 + C_2)} \tag{11}$$

$\mu_Y$, $\mu_{\hat{Y}}$, $\sigma_Y^2$, $\sigma_{\hat{Y}}^2$, and $\sigma_{Y\hat{Y}}$ are local

PSNR, which measures the ratio between the maximum possible signal power and the power of noise affecting the accuracy of the image representation, is defined as follows:

$$PSNR(Y, \hat{Y}) = 20 \cdot \log_{10}\left(\frac{MAX(Y)^2}{\sqrt{MSE(Y, \hat{Y})}}\right). \tag{11}$$

The resolution-scaled Pearson coefficient and resolution-scaled error are calculated using the Nanoj-SQUIRREL (No-GPU version) plugin in ImageJ. In this process, the algorithm uses a low-resolution ground-truth (GT) image of a single organelle and the output image from Deep-DSCM as reference images. The super-resolution reconstructions are convolved with a calculated PSF, which is automatically estimated and subsequently compared to the low-resolution GT images. Because our degradation model operates in three sequential stages, we can generate multi-colour, low-resolution GT images tailored for resolution-scaled calculations by applying only the blurring stage of the degradation model to each channel. The resolution is determined by using single-frame rFRC mapping via the PANELJ plugin for ImageJ. For each dataset, the background intensity and pixel size are configured alongside default parameter settings.

## Acknowledgements
This work has been supported by the National Key R&D Program of China (2021YFF0502900); National Natural Science Foundation of China (T2421003/62127819/62435011); Guangdong Basic and Applied Basic Research Foundation (2024A1515030193/2023A1515010795); Shenzhen Science and Technology Program (JCYJ20220818100202005). We thank Dr. Yue Chen from the Photonics Center of Shenzhen University for her assistance in sample preparation. We also thank Springer Nature for editing and professional language polishing of our manuscript.

## Data availability
The data that support the findings of this study are available from the corresponding author upon reasonable request.

## Code availability
The pytorch codes of Deep-DSCM, including the evaluation of synthetic data and visualization of real data, are publicly available at https://github.com/Chen13d/Deep-DSCM.

## Author contributions

Q.C., L.W. and J.Q. conceived the idea and designed the project. L.L., D.J. and J.Q. supervised the research. Q.C., D.S. and J.L. developed and improved the algorithm. L.W. and X.W. designed and built the microscope. Q.C. and L.W. prepared biological samples, performed experiments and data processing. Q.C., L.W., J.L., D.J. and J.Q. wrote and revised the paper with input from all authors.

**Competing interests**

The authors declare no conflicts of interest.

**Supplementary information**

See in Supplementary materials.

# Figures

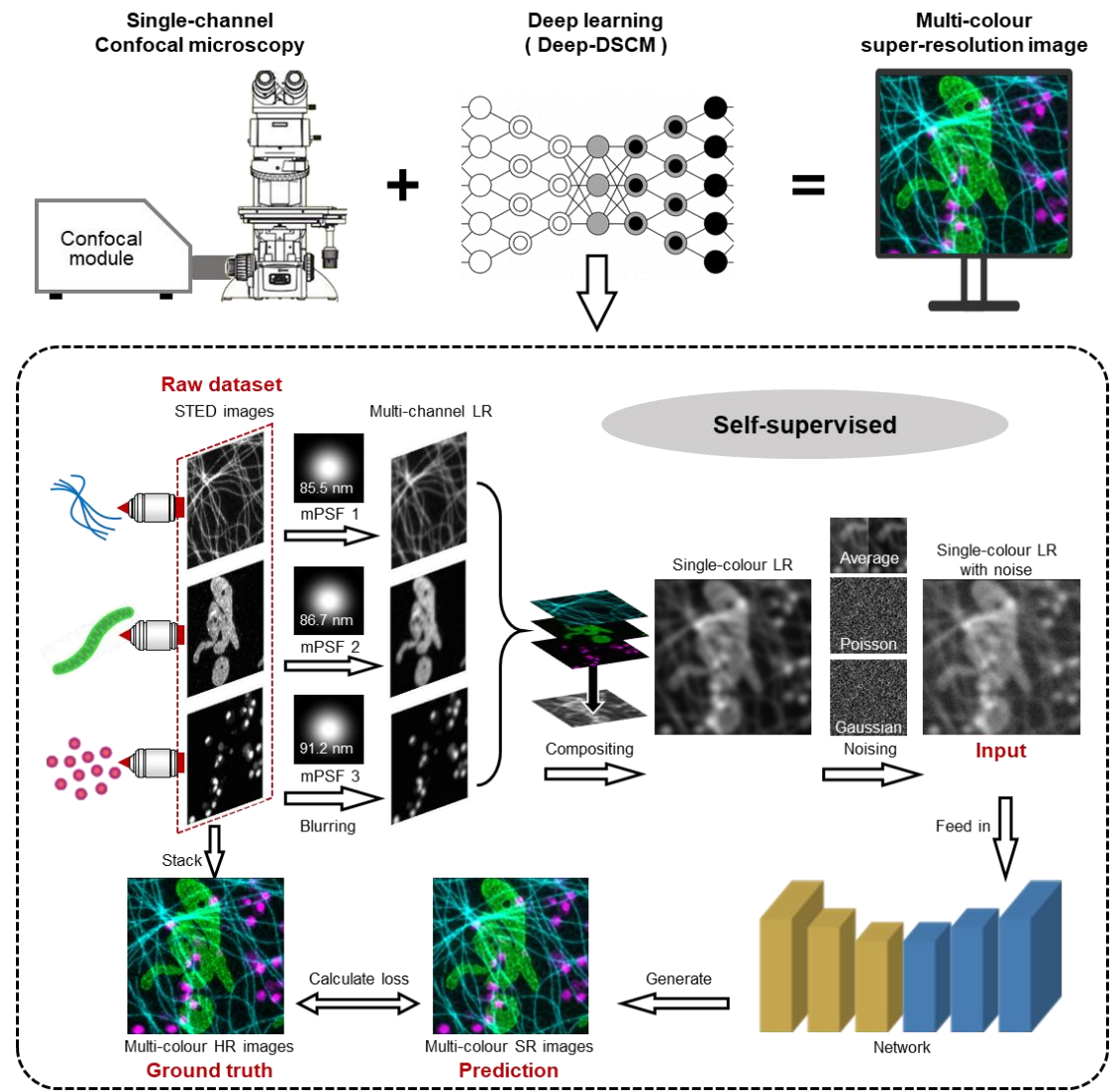

**Fig. 1 Workflow of Deep-DSCM.**

The procedure to obtain a multi-colour super-resolution image. The Deep-DSCM model starts with a single-channel confocal microscope, by that one can acquire single-colour confocal image, then the single-colour confocal image is sent to trained Deep-DSCM model to generate multi-colour super-resolution image. The box includes the self-supervised training process of Deep-DSCM. Both the multi-colour high-resolution images and single-colour low-resolution images with noise are generated from the raw dataset, which contains single-channel STED images from independent cell samples only. The degradation model includes blurring, composition and noising parts, the model generates different mPSFs for each organelle. The noising part includes Poisson, Gaussian noise and frame averaging. The U-Net based Deep-DSCM network takes single-colour low-resolution images as input and generate prediction, to compare with the multi-colour high -resolution images and calculate loss.

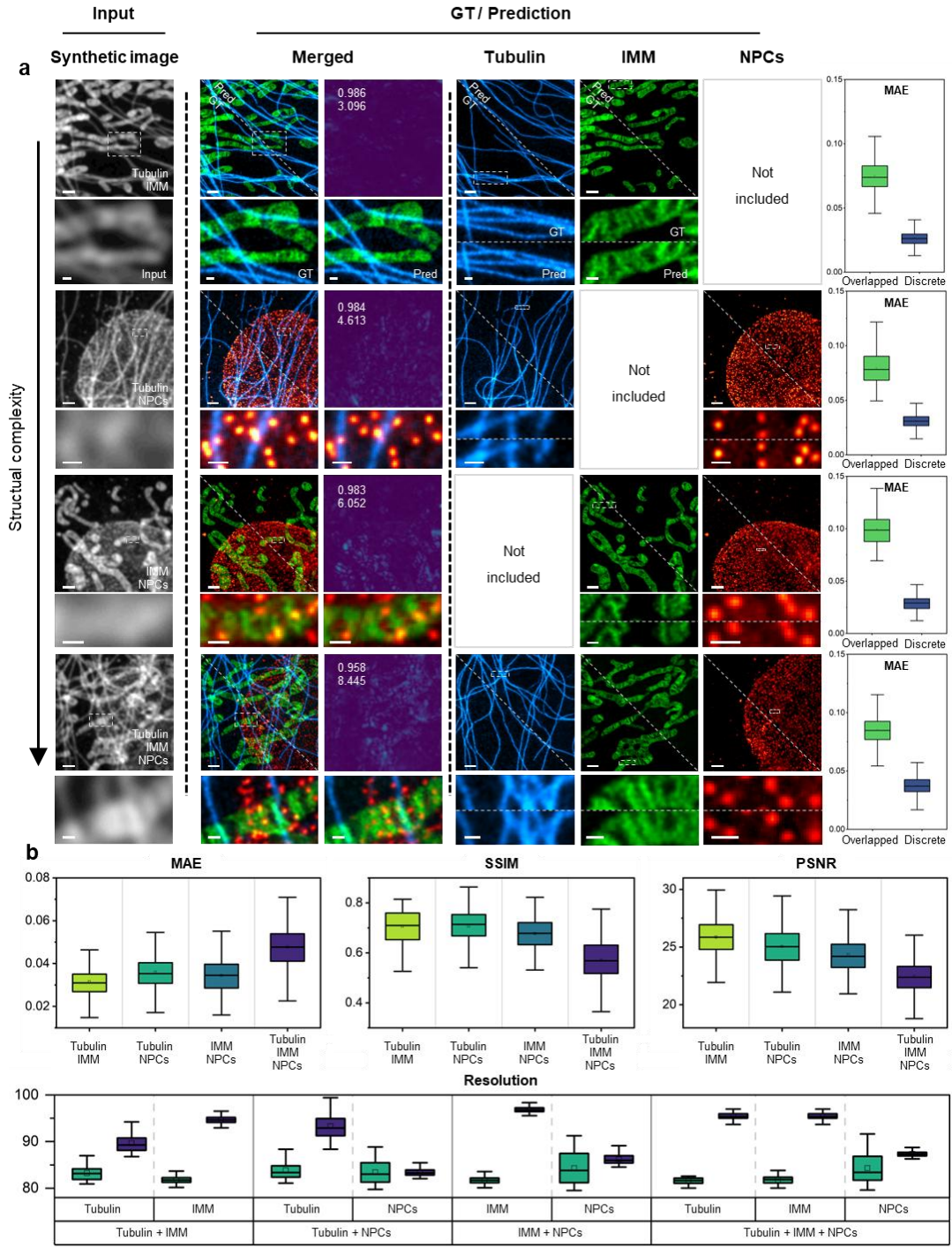

**Fig. 2 Results on synthetic data.**

**a** Results on synthetic datasets containing different organelle compositions: Microtubules (Tubulin)., inner mitochondrial membrane (IMM), and nuclear pore complexes (NPCs). The leftmost column shows an overview along with the input single-colour synthetic images. On the right, multi-colour super-resolution GT and output images for each organelle are displayed in merged form, separated by a dashed line. Corresponding RSE maps, alongside their respective resolution scaled Pearson coefficient (RSP) and resolution scaled error (RSE) values are also presented. Separated frames are presented thereafter. In the rightmost shown the box plot of mean absolute error (MAE) comparison between the overlapped and discrete regions Enlarged views of the GT and output images are shown below each pair for better visualization (dashed box). Scale bar, 1 μm for the full images, 200 nm for the enlarged regions. **b** Statistical evaluation of the synthetic dataset, as shown in b, using metrics such as the MAE structural similarity index matrix (SSIM), peak signal-to-noise ratio (PSNR), and single-frame rFRC mapping. Box plots compare these metrics across the different organelle compositions. The resolution analysis includes distributions from both GT and output images to assess performance.

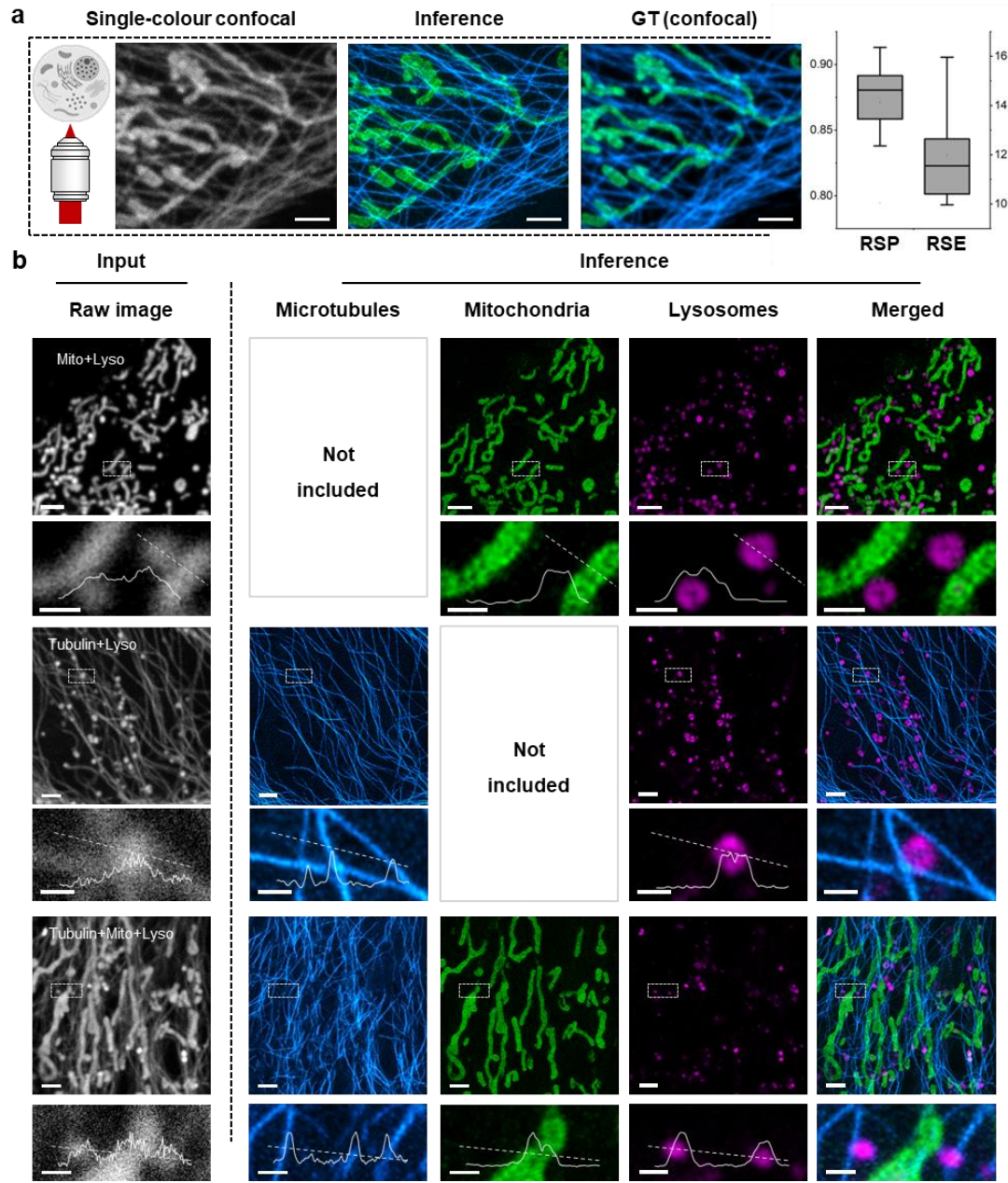

**Fig. 2 Results on synthetic data.**

**a** Results on the staining combination of microtubules–mitochondria, which is imaged in three channels, yielding both input and GT images, the inference image is generated by the trained Deep-DSCM model. The right panel presents the statistical results for RSP and RSE. **b** Results on other staining combinations, including mitochondria–lysosomes, microtubules–lysosomes and microtubules–mitochondria-lysosomes. Each row includes an overview image and the input confocal images on the left, while the decoupled and merged super-resolution images of the organelles are displayed on the right. Below each set of images, magnified views of both the input and inference images are provided with the dash lines and corresponding profiles. Scale bar: 2 μm for whole images, 500 nm for magnified views.

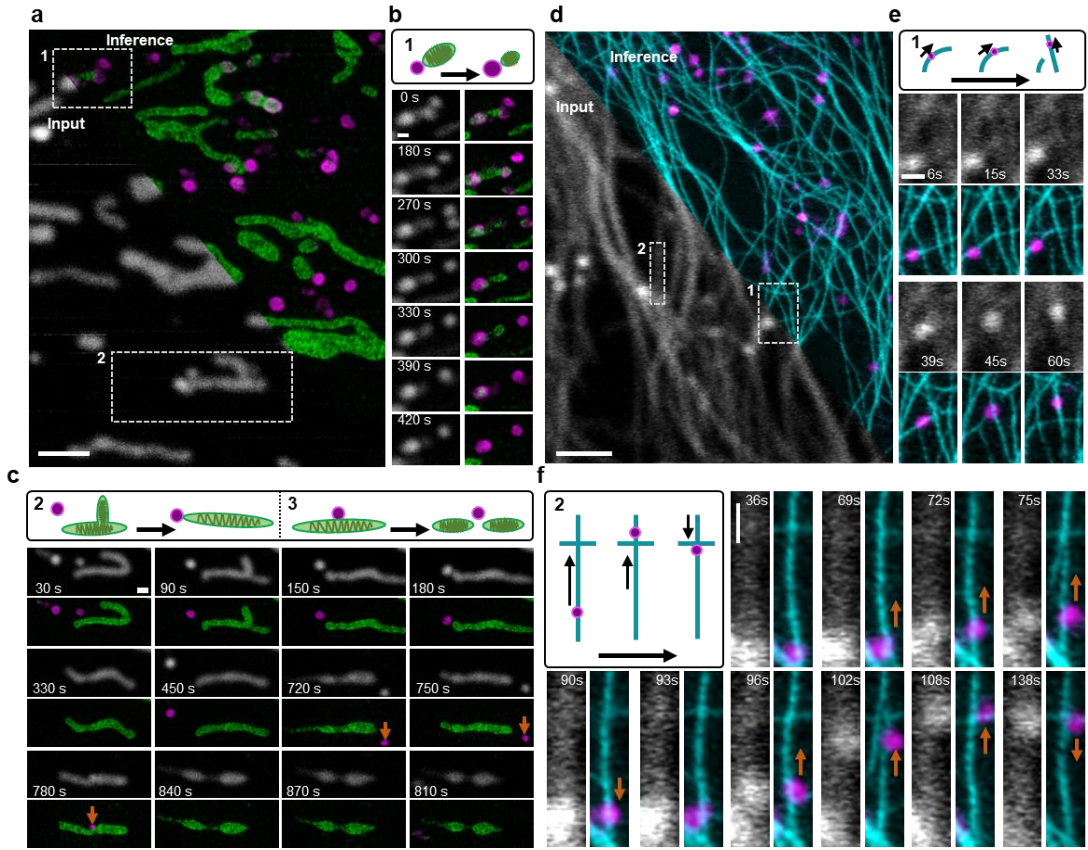

**Fig. 4 Deep-DSCM helps with organelles interaction observations.**

**a**–**c** Observation of mitochondrion-lysosome interaction, with trained Deep-DSCM network. **a** Raw single-colour confocal image and corresponding network inference, the dashed boxes represent region of **b** and **c**. Scale bar, 2 μm. **b** Degradation process between mitochondria and lysosomes, and shown in box 1 in **a**, scale bar: 500 nm. **c** Magnified images of box 2 in **a**, consisting of both "Kiss and run" and fission process of mitochondria and lysosomes. The yellow arrow corresponds to the involved lysosome for fission. Scale bar, 500 nm. **d**–**f** Observation of microtubule-lysosomes movement. d Raw single-colour confocal images and corresponding network inference, the dashed boxes represent region of **e** and **f**. **e** Magnified images of box 1 in **d**, displaying a lysosome switch from one microtubule to another, scale bar: 500 nm. **f** Magnified images of box 2 in **d**, showing a lysosome moving back and forth along the microtubule and microtubule intersection, scale bar: 500 nm.

# Supplementary information

# Self-supervised learning for multiplexing super-resolution confocal microscopy

Qinglin Chen[1,†], Luwei Wang[1,†], Jia Li[1], Dan Shao[2], Xiaoyu Weng[1], Liwei Liu,[1], Dayong Jin[2,3,*] & Junle Qu[1,*]

[1]State Key Laboratory of Radio Frequency Heterogeneous Integration (Shenzhen University) & Key Laboratory of Optoelectronic Devices and Systems, College of Physics and Optoelectronic Engineering, Shenzhen University, Shenzhen 518060, China

[2]Institute for Biomedical Materials and Devices (IBMD), Faculty of Science, University of Technology Sydney, NSW 2007, Australia

[3]Zhejiang Provincial Engineering Research Center for Organelles Diagnostics and Therapy, Eastern Institute of Technology, Ningbo 315200, China

[†]These authors contributed equally: Qinglin Chen, Luwei Wang.

[*]Corresponding author: dayong.jin@eitech.edu.cn; jlqu@szu.edu.cn

# Supplementary Information

In this supplementary material, we describe the network structure, workflow of degradation model and sample preparation configuration, along with supplementary results of Deep-DSCM.

## 1. Network structure

The Deep-DSCM network consists of two U-Net-based structures: a generator and a discriminator. The U-Net [**Error! Reference source not found.**] architecture is used for the generator because of its strong performance in reconstructing both coarse and fine structures. As described in [2], the discriminator, also based on U-Net, produces outputs at two scales: a downsampled encoded output, $D_{enc}(G(X))$, and a decoded output at the original resolution, $D_{dec}(G(X))$. These outputs enable per-pixel evaluation at multiple scales.

The generator follows an encoder-decoder structure, referred to as $G_{enc}$ and $G_{dec}$. In $G_{enc}$ (shown with a light purple background), the input is progressively downsampled through a series of ConvBlock (light orange), which double the feature dimensions at each step (e.g., 64, 128, 256). Downsampling is performed using 2×2 Maxpooling layers (plum). Each ConvBlock contains two sequential operations: a 3×3 Conv2D (stride of 1, padding of 1) followed by batch normalization and LeakyReLU activation (with a slope of 0.1). These MaxPooling layers control the downsampling process between ConvBlocks.

In $G_{dec}$ (blue background), features are upsampled using nearest neighbor interpolation and $1 \times 1$ convolution layers (Cyan) instead of transposed convolution [**Error! Reference source not found.**]. The encoded features are progressively upsampled by halving the feature dimensions at each step (e.g., 512, 256, 128). Skip connections are added between corresponding layers in $G_{enc}$ and $G_{dec}$ through concatenation, allowing fine-grained details to pass efficiently through the network. The generator's final layer, part of the OutputBlock (pale pink), applies ReLU activation, as the dataset only contains non-negatives values.

The discriminator also adopts an encoder-decoder structure, split into $D_{enc}$ and $D_{dec}$ [2]. Like the generator, $D_{enc}$ (light green background) downsamples the input using multiple DownsampleBlocks while increasing the feature dimensions (e.g., 64, 128, 192). Each DownsampleBlock includes a $3 \times 3$ convolution (stride of 1, padding of 1), batch normalization, LeakyReLU activation (slope of 0.1), and $2 \times 2$ average pooling. In $D_{dec}$ (yellow background), features are progressively upsampled using UpsampleBlocks, reducing feature dimensions (e.g., 384, 320, 256) at each step. Each UpsampleBlock contains a $3 \times 3$ convolution (stride of 1, padding of 1), batch normalization, nearest neighbor interpolation, and LeakyReLU activation (slope 0.1). Skip connections are implemented between corresponding encoder and decoder blocks by concatenating the DownsampleBlocks and UpsampleBlocks.

At the end of $D_{enc}$ and $D_{dec}$, a 1×1 convolution layer (stride of 1, padding of 0) generates decision maps, $D_{enc}(G(X))$ and $D_{dec}(G(X))$, which evaluate the fidelity of the generator's output on a per-pixel level.

During forward propagation, these decision maps are compared with the label maps to calculate the adversarial loss. The label maps are set to ones for GT images and zeros for generated images. Combined with MSE loss, this adversarial loss allows us to update the Deep-DSCM network through backpropagation.

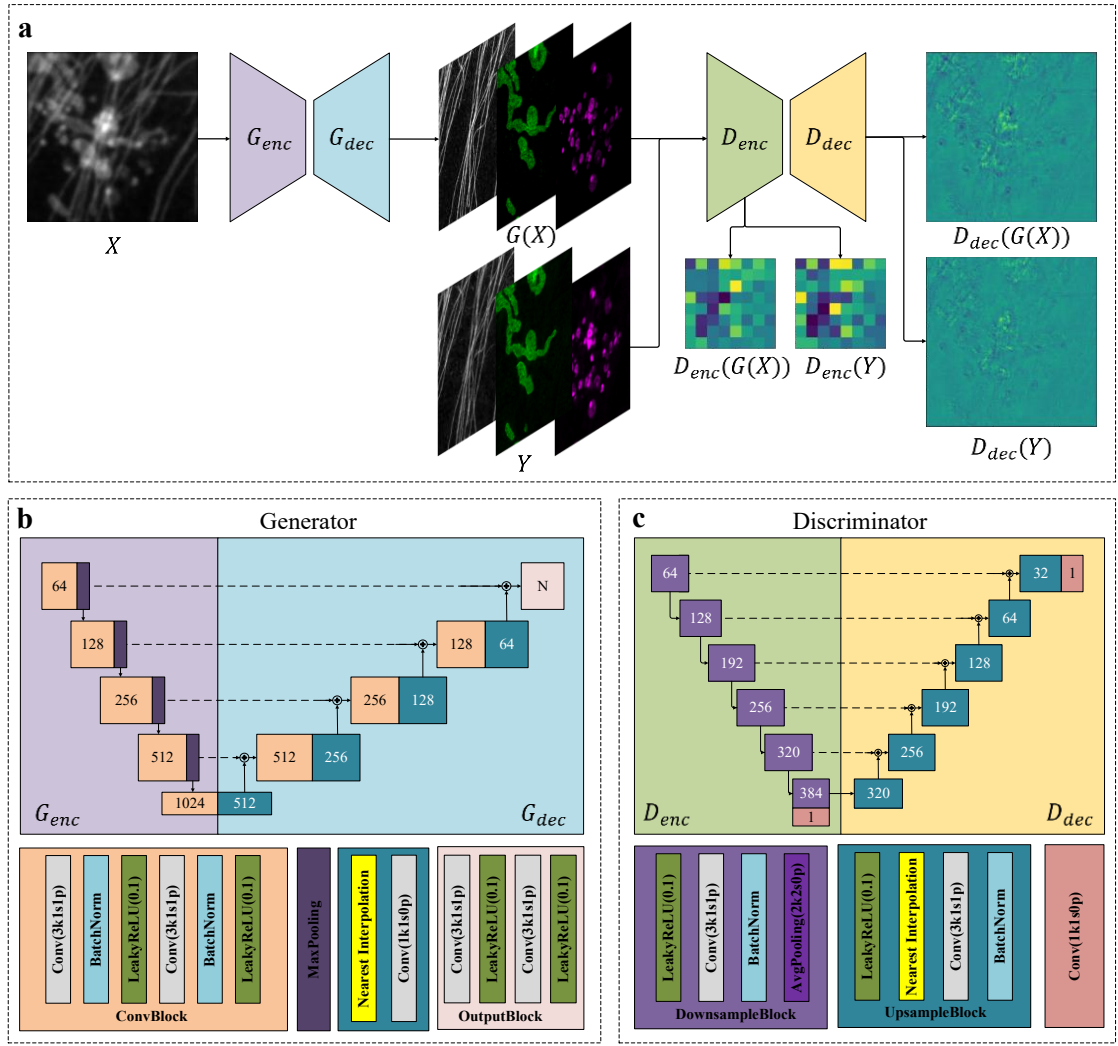

**Fig. S1 Deep-DSCM network structure. a** Overall structure of the Deep-DSCM network, showing its forward process and corresponding outputs. **b** Detailed design of the generator. **c** Detailed design of the discriminator.

## 2. Procedure for the degradation model

As shown in Figure S2, the multi-channel super-resolution image, which serves as the GT for the network, consists of images of multiple organelles, each represented by a separate channel. First, the multi-channel super-resolution image is scaled by different factors and combined into a single channel to create a single-channel super-resolution image. To degrade its resolution, we used methods described in [4].

Specifically, we generate a modulated point spread functions (PSF) using two PSFs estimated from the confocal image and the STED image. The single-channel super-resolution image is then convolved with the modulated PSF, to generate the single-channel low-resolution image.

To replicate real imaging conditions, different levels of noise were added to produce a single-channel low-resolution image with noise. This degraded image was used as input for the network.

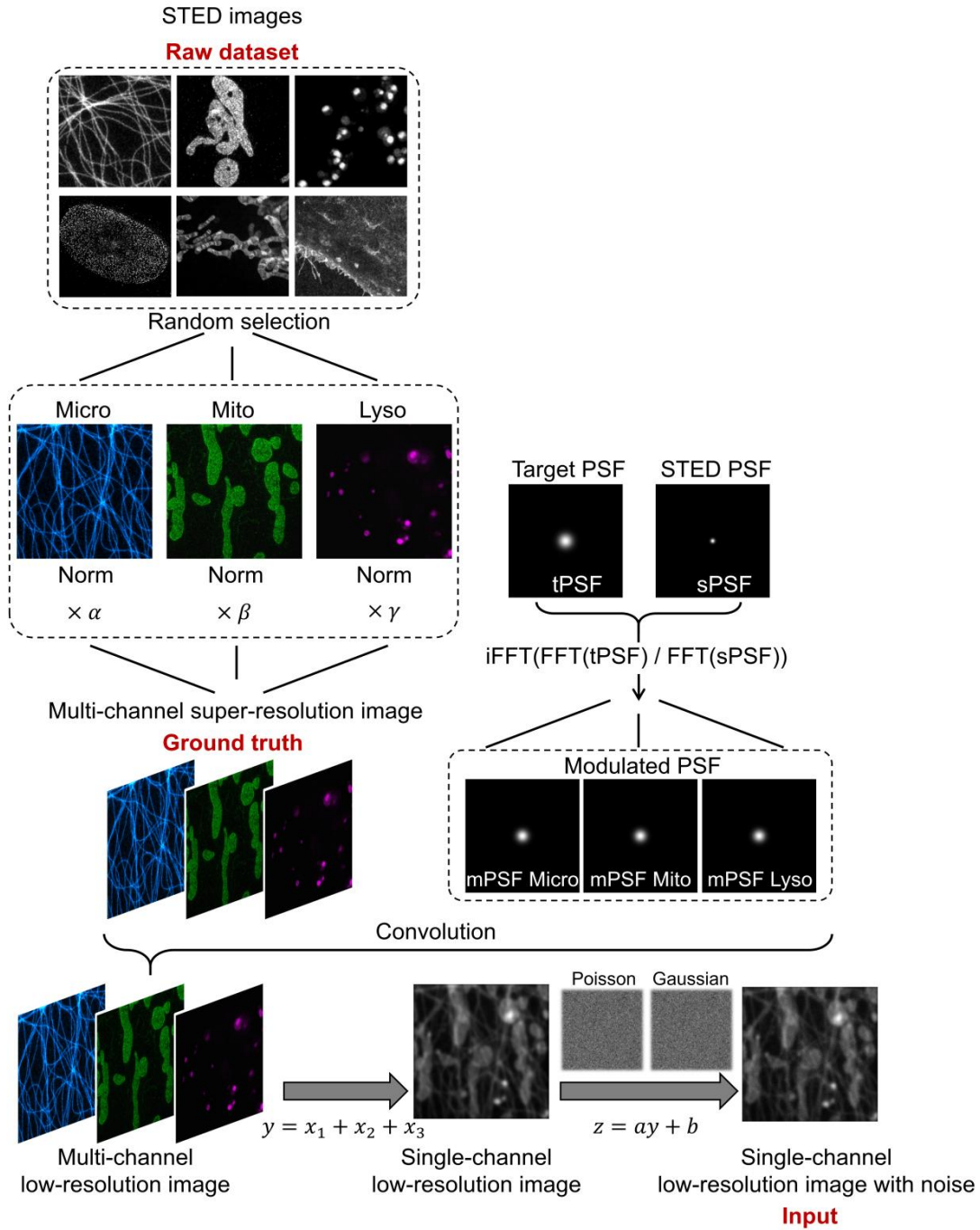

**Fig. S2 Workflow of degradation model.** It illustrates the process of converting a multi-channel super-resolution image into a single-channel, low resolution image.

### 3. Simulation experiments

As shown in Figure S9a, we conducted comparison experiments on simulated data, in which line and ellipse structures were randomly generated, categorized into four independent levels, indicating different structural similarities between components. From level 1 to level 3 the "line" and "ellipse" structures tends to be the same, while in level 4 both structures are line structures. The results are shown in Figure S9b, for level 1 and level 2, the components were well separated and restored since the structures were distinct from each other, the average MAE, SSIM, PCC value had reached over 0.0128, 0.9226 and

0.9936 (the yellow number). In level 3, the structural similarities between components have reached the SSIM value of 0.994, in this case it is not possible for one to distinguish them by examination. However, through learning the small difference between these two structures, the trained model was still able to separate them, but in deteriorated performance, the average MAE, SSIM, PCC value fell to 0.0291, 0.4330 and 0.9430. These results demonstrated that the proposed method is capable even if the structures are quite similar, and revealed the fact that morphology is essential factor for the proposed method. In the case the structures are identically same, for level 4, we generated the structures only using lines, for this time the train model can not separate them, instead, it gave ambiguous distribution of two structures, making the results not reliable anymore, the average MAE, SSIM, PCC is 0.1207, 0.1233 and 0.5267.

When structures are morphologically identical, simply learning the mapping of input and GT is not helpful, other information is required in this situation. Fluorescence imaging is influenced by many factors like staining parameters, system parameters and component properties, causing different intensity for components. Thus, we simulated another condition that the two structures involved are distributed in different value range (145~200 and 200~255), the results are shown in Figure S9c. In level 1 and 2 in which the structures are distinguishable from each other, the results on intensity-different dataset kept close to the intensity-same dataset, when the morphology of each kind tends to be same (level 3, SSIM=0.994), the network still gave separate and clean results, while slight errors occur in Figure 9b. As the two structures become identical (level 4), the intensity-different dataset is still able to recover the lines into dual channels. The above results indicate that in our method, both morphology and intensity feature are important contributors to the reconstructions accuracy.

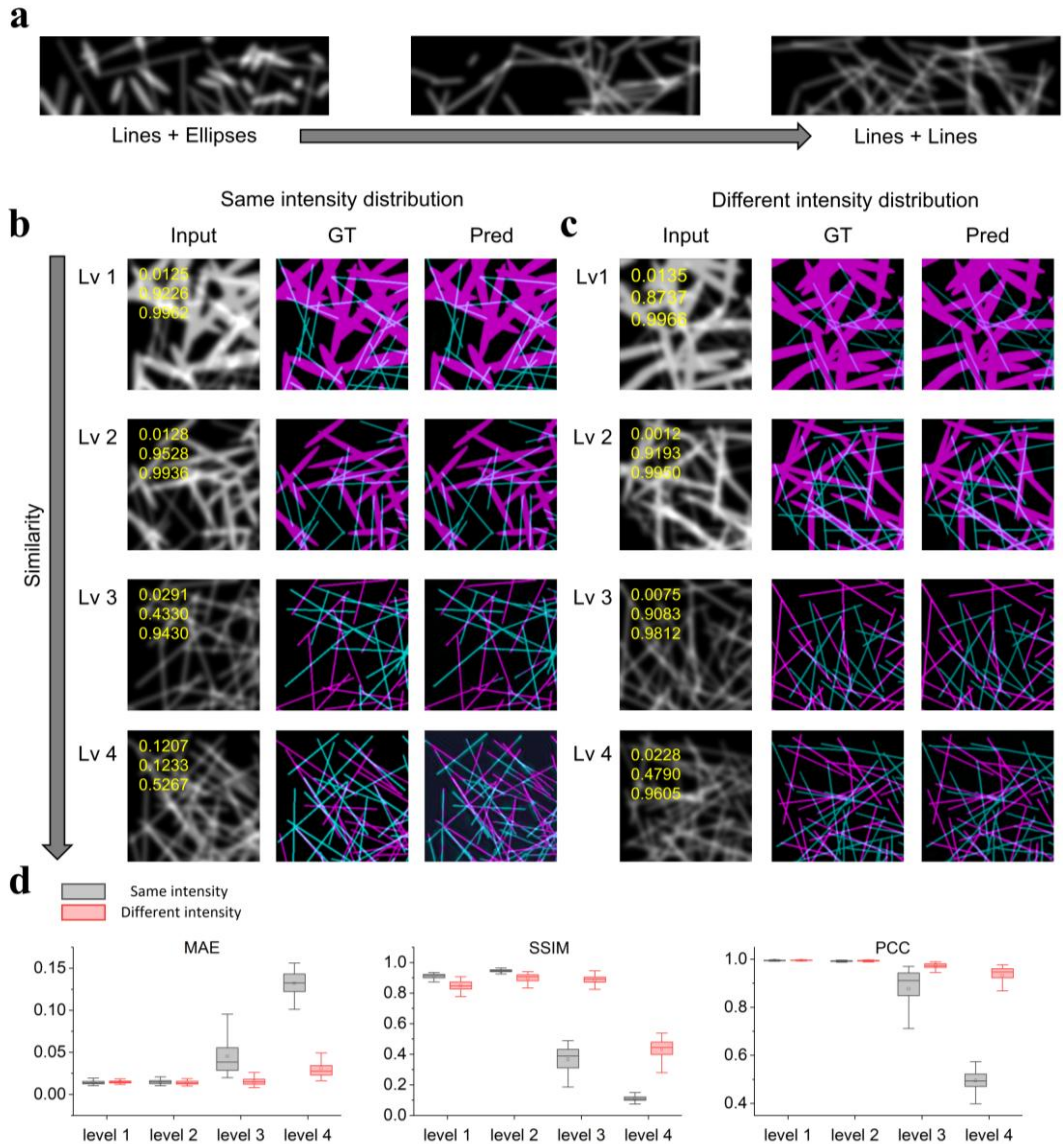

**Fig. S3 Experiments on simulated data. a** A skim on the simulated data, which transits from lines+ellipse to lines+lines. **b** The results of same-intensity dataset, the yellow values indicate MAE, SSIM and PCC by rows for each level, input, GT and prediction were listed by columns. **c** The results of different-intensity dataset. Input, GT and prediction listed by columns. **d** The quantification of MAE, SSIM and PCC of **b** and **c**.

## 4. The sample preparation configurations

This note outlines the preparation methods for samples used in the study. Table S1 details the preparation configurations for living samples, including the sample names, associated cell lines, fluorescent dyes, and incubation times. Both single-stained and multi-stained samples are included.

As shown in Table S2, each organelle was individually stained and imaged using a STED microscope under different conditions. The raw pixel size was adjusted to meet the requirements of the Nyquist Theorem. Line accumulations differed depending on the organelle and the fluorescent dyes used.

Table S3 presents samples containing various compositions of microtubules, mitochondria, and lysosomes. During sample preparation, the organelles were stained with fluorescent dyes that had overlapping emission wavelengths. The samples were then imaged using a single channel confocal microscope. Accumulations were not applied to the living cell samples (the last three columns) because the cells were in motion during imaging.

Table. S1 Preparation configurations for different living cell samples

| Sample | Inner mitochondria membrane | Lysosomes | Mitochondria, lysosomes | Microtubules, lysosomes | Microtubules, mitochondria, lysosomes |
|---|---|---|---|---|---|
| Cell line | HeLa, MDA | MDA | HeLa | HeLa | HeLa |
| Dyes | PK Mito Orange | LysoBrite™ NIR | MitoTracker™ Deep Red FM , LysoBrite™ NIR | Tubulin Tracker™ Deep Red , LysoBrite™ NIR | Tubulin Tracker™ Deep Red , MitoTracker™ Deep Red FM , LysoBrite™ NIR |
| Incubation time | 15 min | 15 min | 30 min | 1 h | 15 min for lysosomes, 30 min for mitochondria outer membrane, 1 h for microtubules |

Table. S2 Setup of different training samples

| Sample | NPCs | Inner mitochondria membrane | Plasma membrane | Microtubules | Mitochondria | Lysosomes |
|---|---|---|---|---|---|---|
| Imaging platform | Facility line | STEDYCON | STEDYCON | STEDYCON | STEDYCON | STEDYCON |
| Objective | 60X, NA1.42 | 100X, NA1.45 | 100X, NA1.45 | 100X, NA1.45 | 100X, NA1.45 | 100X, NA1.45 |
| Excitation | 640 nm | 561 nm | 561 nm | 640 nm | 640 nm | 640 nm |
| Depletion | 775 nm | 775 nm | 775 nm | 775 nm | 775 nm | 775 nm |
| Fluorescence dyes | STAR RED | PK Mito Orange | PK Mito Orange | STAR RED | STAR RED | LysoBrite NIR |
| Line accumulation | 5 | 3 | 3 | 5 | 5 | 3 |
| Raw pixel size | 30 nm | 20 nm | 20 nm | 20 nm | 20 nm | 30 nm |
| Image size | Random | Random | Random | 1000*1000 | 1000*1000 | Random |

| Frames | 104 | 115 | 236 | 224 | 196 | 70 |

Table. S3 Setup of different testing samples

| Sample | Microtubules, Mitochondria | Microtubules, Lysosomes | Mitochondria, Lysosomes | Microtubules, Mitochondria, Lysosomes |
|---|---|---|---|---|
| Imaging platform | SP8, Leica | Home built | Home built | Home built |
| Objective | 100X, NA1.4 | 100X, NA1.4 | 100X, NA1.4 | 100X, NA1.4 |
| Excitation | 638 nm | 635 nm | 635 nm | 635 nm |
| Detection | 660–730 nm | 667 ± 15 nm | 667 ± 15 nm | 667 ± 15 nm |
| Fluorescence dyes | STAR RED, STAR RED | Tubulin Tracker™ Deep Red, LysoBrite™ NIR | MitoTracker™ Deep Red FM, LysoBrite™ NIR | Tubulin Tracker™ Deep Red, MitoTracker™ Deep Red FM, LysoBrite™ NIR |
| Line accumulation | 5 | - | - | - |
| Pixel size | 22.7 nm | 13 - 26 nm | 13 - 26 nm | 13 - 26 nm |
| Image size | 1024*1024 | 1024*1024 | 1024*1024 | 1024*1024 |

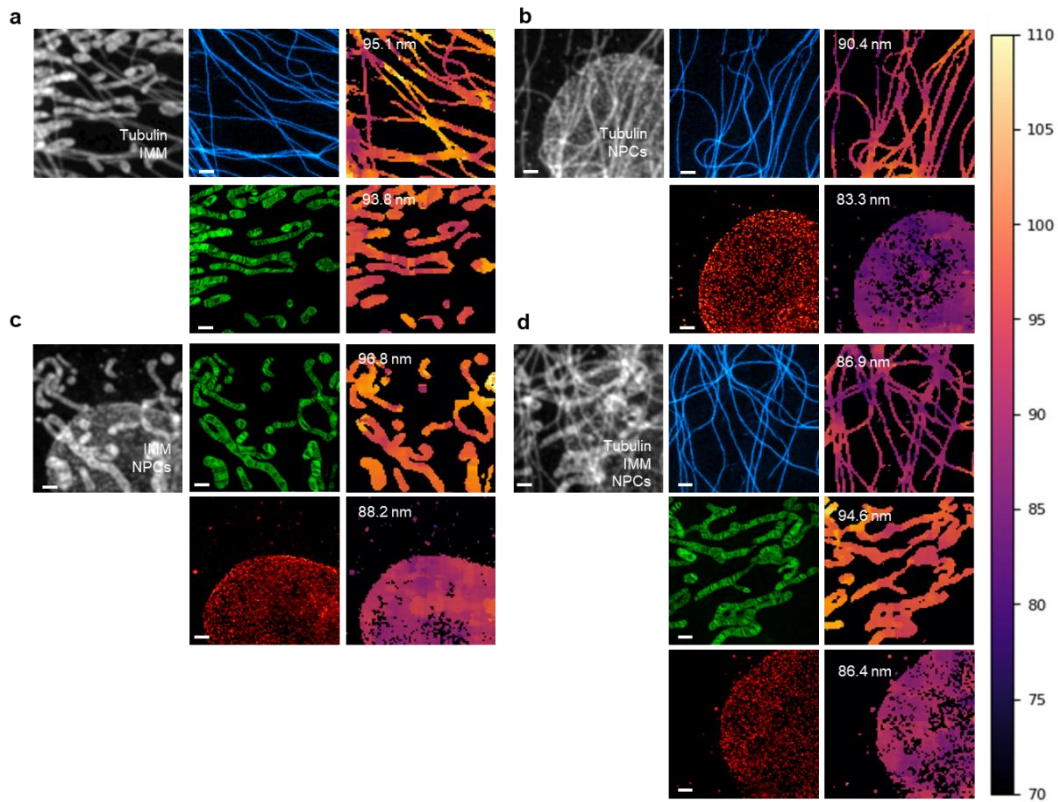

**Fig. S4 Resolution maps of synthetic data for Microtubules, NPCs and IMM.** Composition of **a** Microtubules and IMM, **b** Microtubules and NPCs, **c** IMM and NPCs, **d** Microtubules, IMM and NPCs. The colour bar is shown on the right. Scale bar: 1 μm. For each composition, the input, GT/output, mean resolution values, and their corresponding resolution maps are shown in sequence.

## 5. Deep-DSCM is data-efficient on small amount of data

This note showcases the data efficiency advantage of our method. We prepared the data in small amount (20 raw STED sets for raw dataset) of microtubules-mitochondria composition. By controlling the amount of the data pairs generated by the degradation model, we performed experiments on three different groups, different in the number of data pairs (20, 1000, and 300,000), the last group is in 300,000 pairs because the number of iterations in 300,000 for 1000 epochs, the maximum number of possible composition between organelles in 20 images of 1000*1000 pixel is 57,969,645,888,400 (given the patch size of 384*384 pixel). The results are shown in Figure S4 a, b, for the set of 20 samples, the training converged quickly, while performing poor on validation dataset, the generated channel is still tangled, exhibiting severe structure overflow between channels. As the number of samples increases (1000 samples), the performance on validation dataset is much better, the generated microtubules and mitochondria images are intact, suggesting the advantage of the random selection process in the degradation model. In addition, in 300,000 samples set, the result performance is not proved to be better, while the validation curve is more stable than 1000 samples set. This experiment indicates that the proposed method, the training pairs derive from raw datasets, though the available number of raw dataset is limited, the sample spaces can be augmented by randomly selecting the organelle images and compositing them, simulating different subcellular conditions.

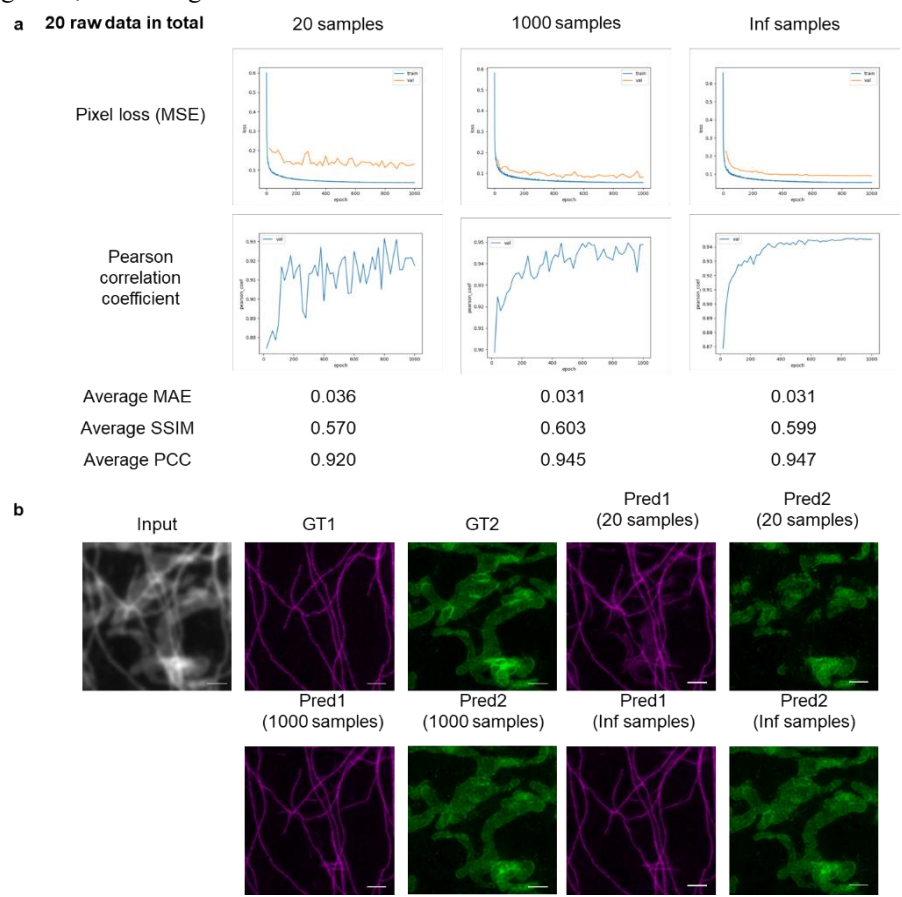

**Fig. S5 Comparison of Deep-DSCM performance on tiny dataset. a** Network performance on different samples configuration, including the training and validation curve, and statistic values. **b** visualization of results of different samples configurations. Scale bar: 1 μm.

## Evaluation of real data on single-structure images

This note demonstrates evaluation results of Deep-DSCM method on real-world single-structure images. Only data from fixed samples was involved due to incomplete data of motive living cell organelles, the fixed sample of microtubules, mitochondria and NPCs (nucleus pore complexes) are shown. The models were trained by the simplified degradation model (only blurring and noising parts were conducted). We applied MAE (mean absolute error), SSIM (structural similarity index matrix) and PCC (Pearson correlation coefficient) to evaluate the precision of predictions. The results are listed in both images and tables. The results show that the synthetic data trained models can have good performance on the real-world data by the simulation of the imaging condition, they generate good results on the microtubules and mitochondria with lower performance on NPCs, due to the *mPSF* in this work does not include depth information, which appears in NPCs' bumped distribution.

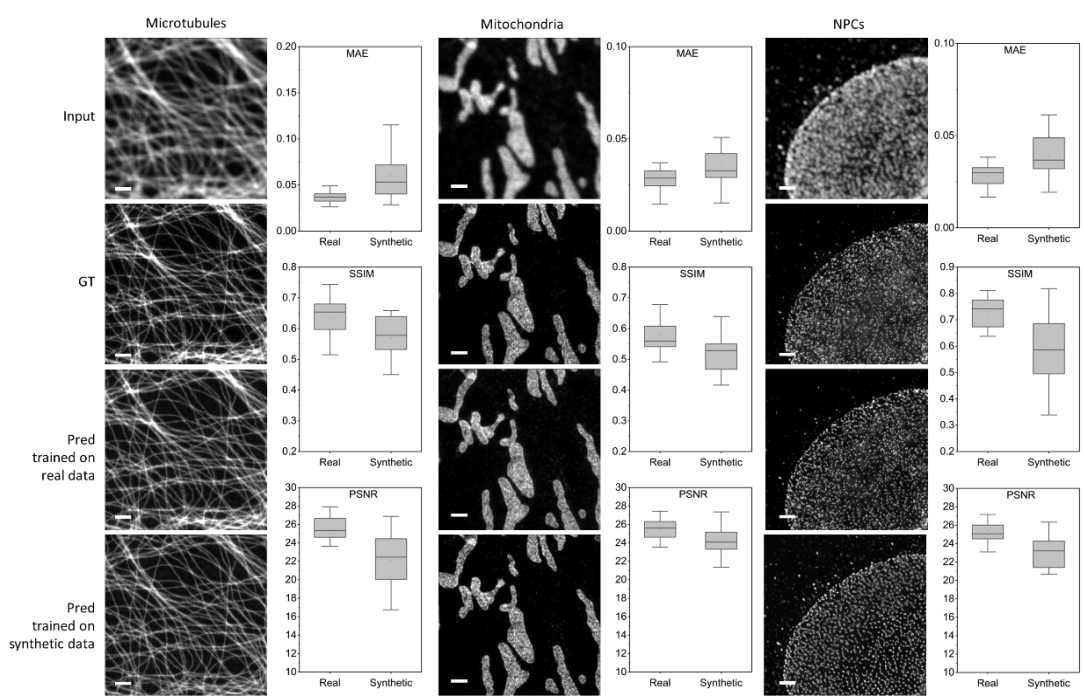

**Fig. S6 Evaluation results of single-structure real data.** The results of different organelle structures are shown in separate rows. For each row, the input, GT, and output are shown. Scale bar: 1 µm.

## Parameters of degradation model affects its performance on confocal images

As described in Methods, the formula for the noising part is:

$$\hat{I}(x,y) = \frac{1}{N}\left[I(x,y) + \frac{Poisson(\alpha \cdot I(x,y))}{\alpha} + N(0,\sigma^2)\right].$$

In this section, we compare the results on real-world single-channel confocal data using different noise parameters ($\alpha$) and target resolutions during training. The parameters N and were set as 4 since the raw data was captured in 4 line accumulations, and $\sigma$ was set to 1. As Figure S7 shows, synthetic data trained network-generated results visually appear to be similar among the parameter combinations, only a slight variation appears, (see in statistical results in Figure S7b)

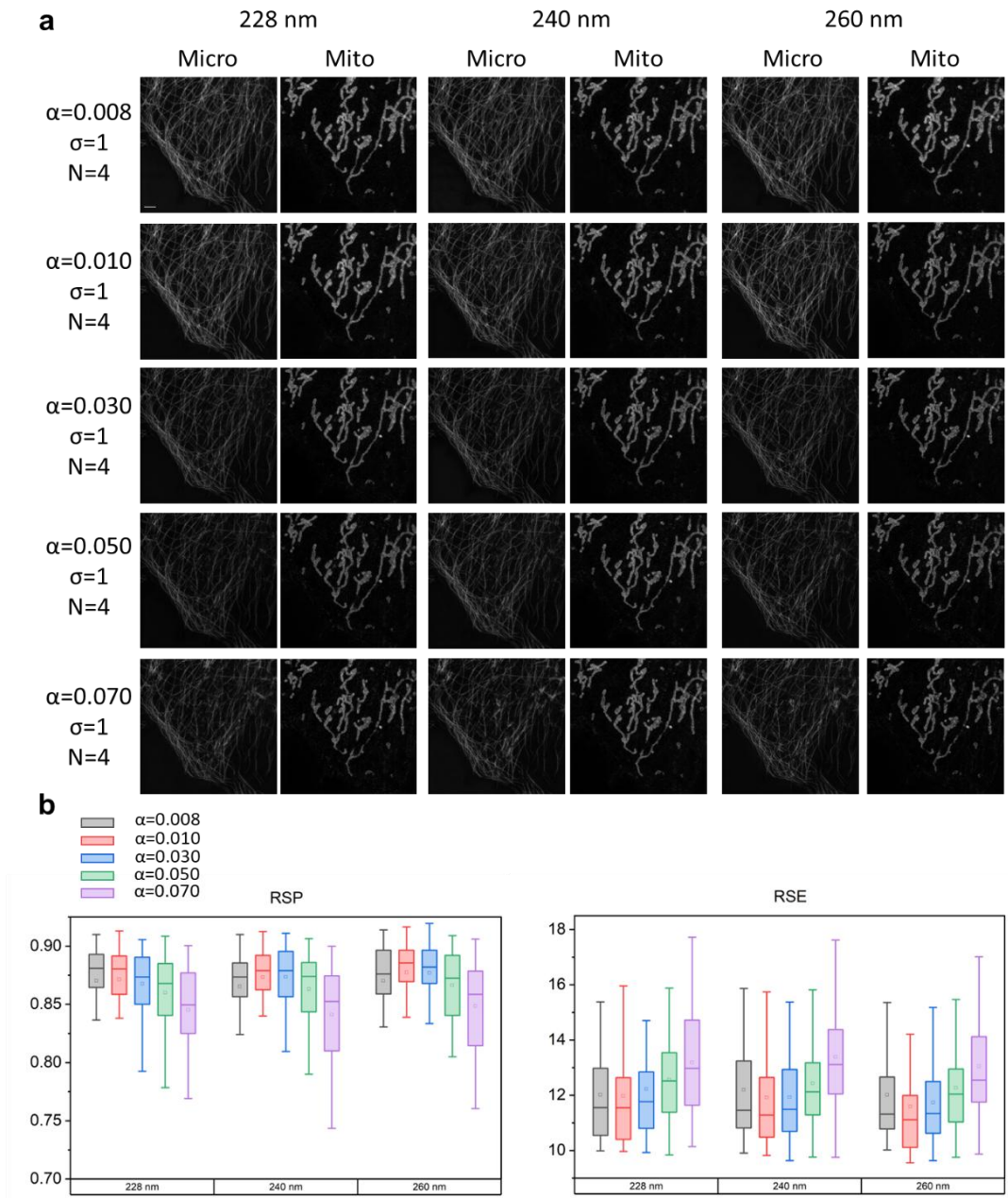

**Fig. S7 Comparison of different parameters on real-world single-channel confocal data**

**a** Comparison of different parameters, scale bar: 2 μm. **b** Statistical results of different parameters.

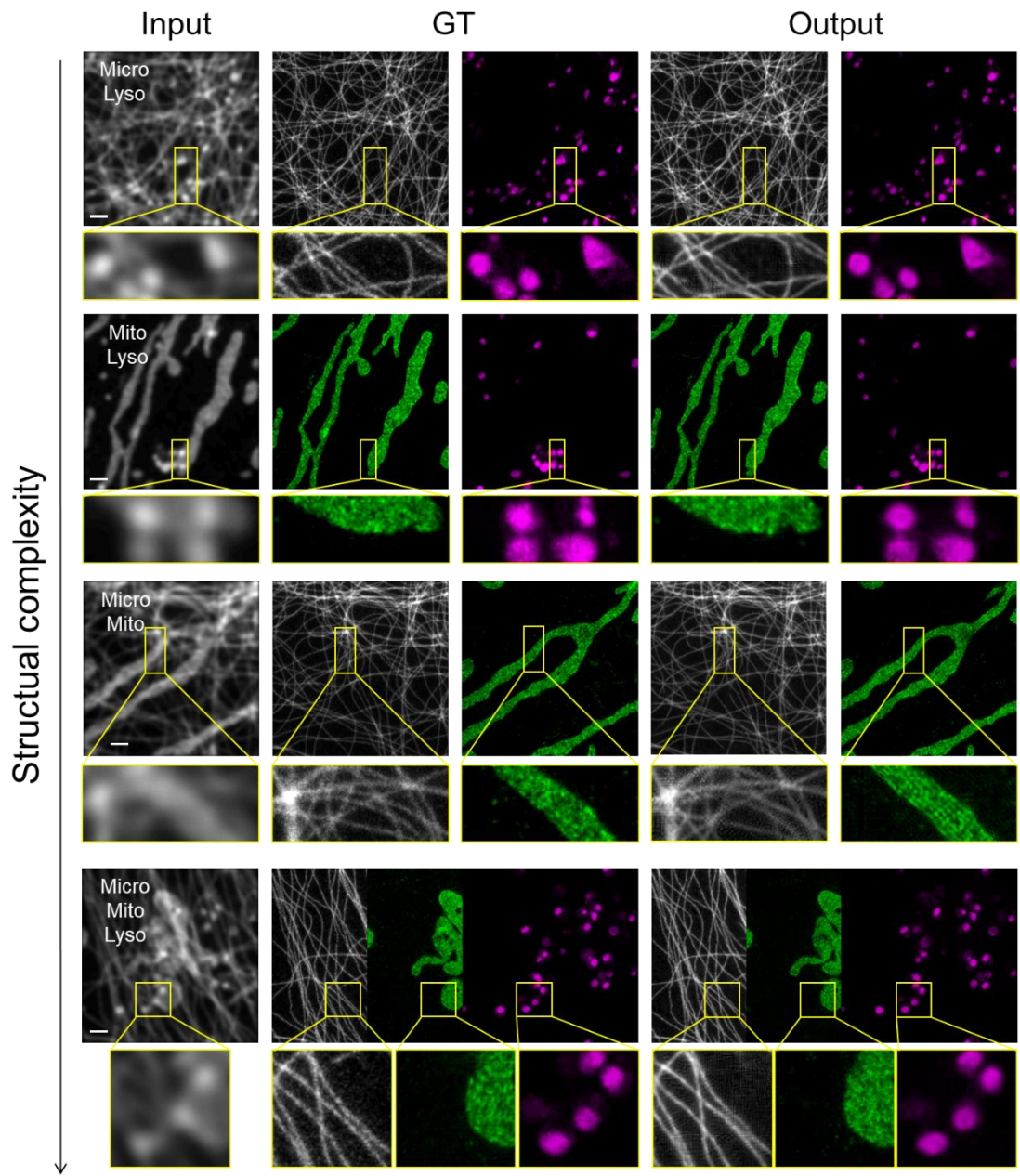

**Fig. S8 Results of synthetic data for microtubules, mitochondria and lysosomes.** The results of different compositions are shown in separate rows, arranged by increasing structural complexity. For each composition, the input, GT, and output images are displayed in order, with magnified views below. Scale bar: 1 μm.

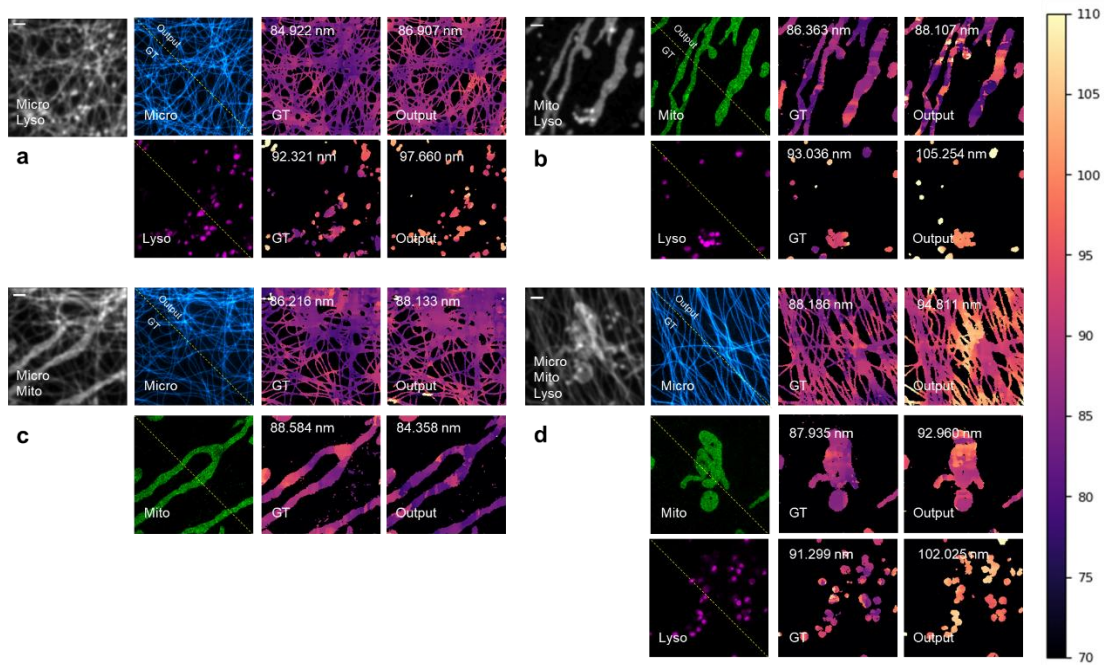

**Fig. S9 Resolution maps of synthetic data for microtubules, mitochondria, and lysosomes.** Composition of **a** microtubules and lysosomes, **b** mitochondria and lysosomes, **c** microtubules and mitochondria, **d** microtubules, mitochondria, and lysosomes. Scale bar: 1 μm. For each composition, the input images, GT images, output images, and resolution maps for both the GT and output images are displayed sequentially. The resolution value is indicated at the top of each resolution map, with the colour bar range set between 70 nm and 110 nm.

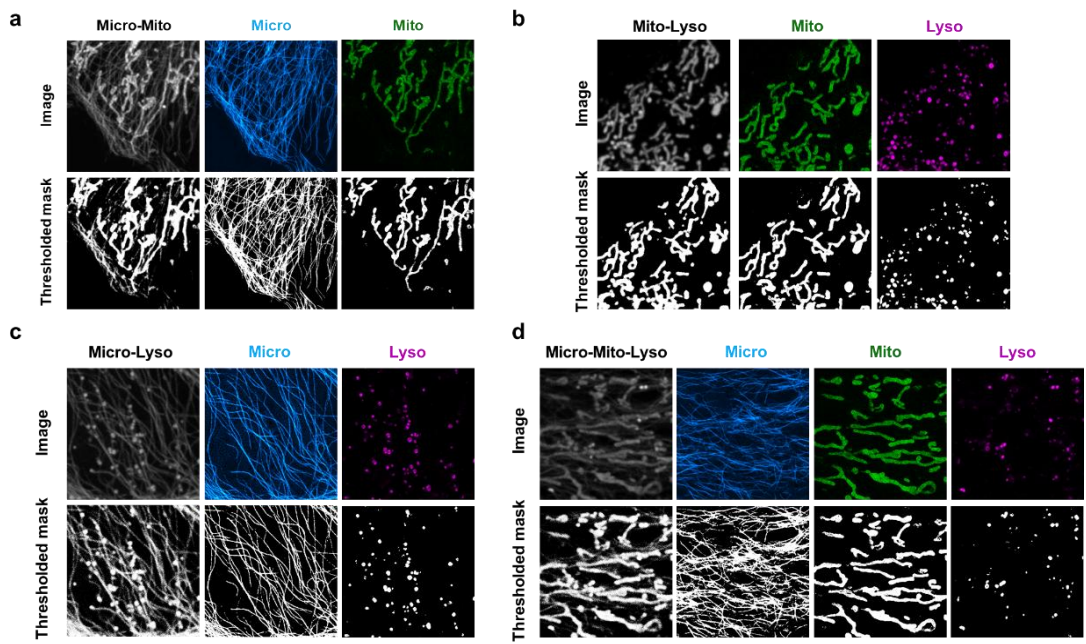

**Fig. S10 Thresholded results in the real dataset.** Thresholding was performed on staining combinations: **a** microtubules and mitochondria, **b** mitochondria and lysosomes, **c** microtubules and lysosomes, **d** microtubules, mitochondria and lysosomes. The operation was conducted by binary function of ImageJ.

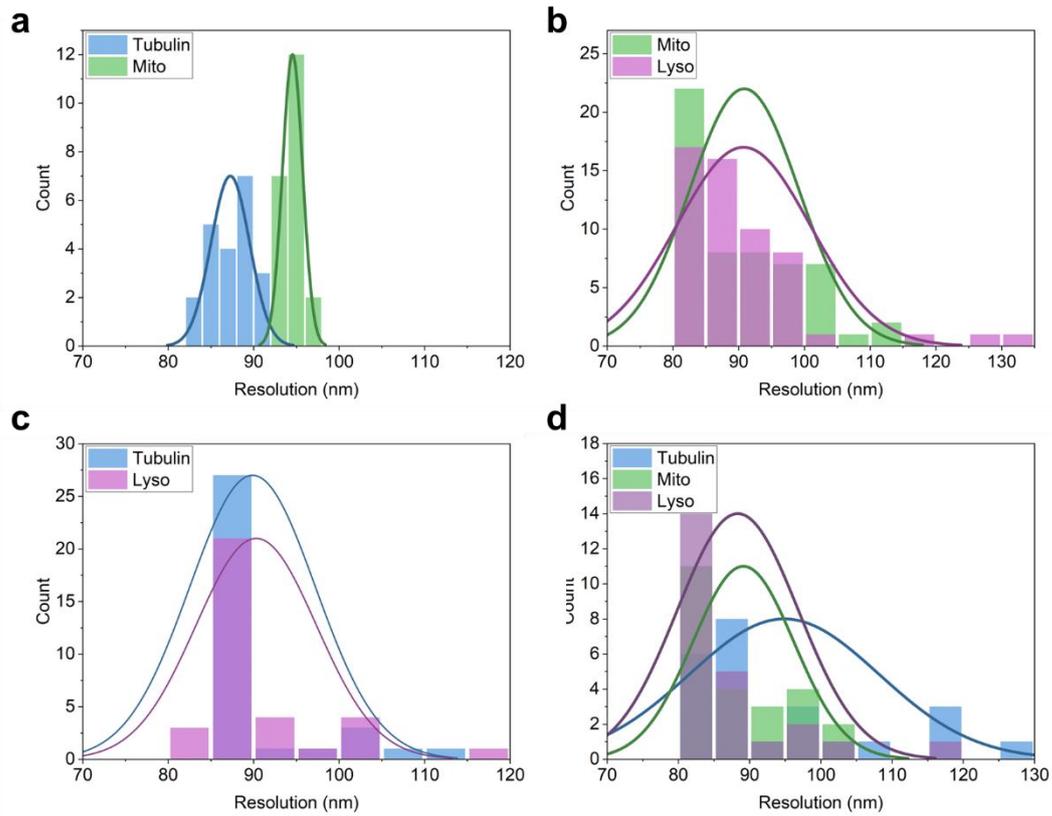

**Fig. S11 Resolution distribution in the real dataset.** The panels present resolution distributions in columns alongside their Gaussian fitting curves for various structural combinations: **a** microtubules and mitochondria, **b** mitochondria and lysosomes, **c** microtubules and lysosomes, and **d** microtubules, mitochondria, and lysosomes. Resolution measurements were performed on the real dataset using single-frame rFRC analysis for all staining combinations, followed by Gaussian fitting of the results.

## Supplementary Bibliography